\documentclass[dvips]{acta}
\usepackage{supertabular,lscape,epsfig}
\usepackage{amssymb}
\usepackage{amsmath}
\usepackage{multirow}
\usepackage{float}
\mathchardef\mhyphen="2D
\usepackage{placeins}
\DeclareSymbolFont{ppa}{OT1}{ppl}{m}{it}
\DeclareMathSymbol{\vv}{\mathalpha}{ppa}{'166}

\SetPages{0}{0}

\SetVol{74}{2024}

\begin{document}
\newcommand\pvalue{\mathop{p\mhyphen {\rm value}}}
\newcommand{\TabApp}[2]{\begin{center}\parbox[t]{#1}{\centerline{
  {\bf Appendix}}
  \vskip2mm
  \centerline{\small {\spaceskip 2pt plus 1pt minus 1pt T a b l e}
  \refstepcounter{table}\thetable}
  \vskip2mm
  \centerline{\footnotesize #2}}
  \vskip3mm
\end{center}}

\newcommand{\TabCapp}[2]{\begin{center}\parbox[t]{#1}{\centerline{
  \small {\spaceskip 2pt plus 1pt minus 1pt T a b l e}
  \refstepcounter{table}\thetable}
  \vskip2mm
  \centerline{\footnotesize #2}}
  \vskip3mm
\end{center}}

\newcommand{\TTabCap}[3]{\begin{center}\parbox[t]{#1}{\centerline{
  \small {\spaceskip 2pt plus 1pt minus 1pt T a b l e}
  \refstepcounter{table}\thetable}
  \vskip2mm
  \centerline{\footnotesize #2}
  \centerline{\footnotesize #3}}
  \vskip1mm
\end{center}}

\newcommand{\MakeTableH}[4]{\begin{table}[H]\TabCap{#2}{#3}
  \begin{center} \TableFont \begin{tabular}{#1} #4 
  \end{tabular}\end{center}\end{table}}

\newcommand{\MakeTableApp}[4]{\begin{table}[p]\TabApp{#2}{#3}
  \begin{center} \TableFont \begin{tabular}{#1} #4 
  \end{tabular}\end{center}\end{table}}

\newcommand{\MakeTableSepp}[4]{\begin{table}[p]\TabCapp{#2}{#3}
  \begin{center} \TableFont \begin{tabular}{#1} #4 
  \end{tabular}\end{center}\end{table}}

\newcommand{\MakeTableee}[4]{\begin{table}[htb]\TabCapp{#2}{#3}
  \begin{center} \TableFont \begin{tabular}{#1} #4
  \end{tabular}\end{center}\end{table}}

\newcommand{\MakeTablee}[5]{\begin{table}[htb]\TTabCap{#2}{#3}{#4}
  \begin{center} \TableFont \begin{tabular}{#1} #5 
  \end{tabular}\end{center}\end{table}}


\newcommand{\MakeTableHH}[4]{\begin{table}[H]\TabCapp{#2}{#3}
  \begin{center} \TableFont \begin{tabular}{#1} #4 
  \end{tabular}\end{center}\end{table}}

\newfont{\bb}{ptmbi8t at 12pt}
\newfont{\bbb}{cmbxti10}
\newfont{\bbbb}{cmbxti10 at 9pt}
\newcommand{\uprule}{\rule{0pt}{2.5ex}}
\newcommand{\douprule}{\rule[-2ex]{0pt}{4.5ex}}
\newcommand{\dorule}{\rule[-2ex]{0pt}{2ex}}
\def\thefootnote{\fnsymbol{footnote}}
\begin{Titlepage}

\Title{The OGLE Collection of Variable Stars: \\
Over 18\,000 Rotating Variables toward the Galactic Bulge}

\Author{
P.~~I~w~a~n~e~k$^1$,~~
I.~~S~o~s~z~y~\'n~s~k~i$^1$,~~
K.~~S~t~\k{e}~p~i~e~\'n$^1$,~~
S.~~K~o~z~{\l}~o~w~s~k~i$^1$,~~
J.~~S~k~o~w~r~o~n$^1$ \\
A.~~U~d~a~l~s~k~i$^1$,~~ 
M.\,K.~~S~z~y~m~a~\'n~s~k~i$^1$,~~
M.~~W~r~o~n~a$^{4,1}$,~~
P.~~P~i~e~t~r~u~k~o~w~i~c~z$^1$ \\
R.~~P~o~l~e~s~k~i$^{1}$~~ 
P.~~M~r~\'o~z$^1$,~~
K.~~U~l~a~c~z~y~k$^{2,1}$~~
D.\,M.~~S~k~o~w~r~o~n$^1$,\\
M.~~G~r~o~m~a~d~z~k~i$^1$,~~
K.~~R~y~b~i~c~k~i$^{3,1}$,~~
M.\,J.~~M~r~\'o~z$^1$~~and~~
M.~~R~a~t~a~j~c~z~a~k$^1$~~ 
}
{$^1$Astronomical Observatory, University of Warsaw, Al. Ujazdowskie 4,\\ 00-478 Warsaw, Poland\\
$^2$Department of Physics, University of Warwick, Gibbet Hill Road, Coventry,\\ CV4 7AL, UK \\
$^3$Department of Particle Physics and Astrophysics, Weizmann Institute of Science, Rehovot 76100, Israel \\
$^4$Villanova University, Department of Astrophysics and Planetary Sciences,\\ 800 Lancaster Ave., Villanova, PA 19085, USA}
\Received{June 12, 2024}
\end{Titlepage}

\Abstract{Stellar rotation, a key factor influencing stellar structure and
  evolution, also drives magnetic activity, which is manifested as spots or
  flares on stellar surface. Here, we present a collection of 18\,443
  rotating variables located toward the Galactic bulge, identified in the
  photometric database of the Optical Gravitational Lensing Experiment
  (OGLE) project. These stars exhibit distinct magnetic activity, including
  starspots and flares. With this collection, we provide long-term,
  time-series photometry in Cousins {\it I}- and Johnson {\it V}-band
  obtained by OGLE since 1997, and basic observational parameters, \ie
  equatorial coordinates, rotation periods, mean brightness, and brightness
  amplitudes in both bands. This is a unique dataset for studying stellar
  magnetic activity, including activity cycles.}{Stellar activity
  (1580), Starspots (1572), Stellar flares (1603), Flare stars (540)}

\Section{Introduction}
Among the various phenomena observed in stars, rotation stands out as a
fundamental aspect shaping their structure (Rivinius \etal 2013, Aerts
\etal 2019) and evolution (Maeder and Meynet 2000). It is also responsible
for the formation of magnetic fields through the dynamo mechanism (Parker
1955, Charbonneau 2013) and the inherently associated magnetic activity
manifested in spots or flares (Irwin \etal 2009, Roettenbacher \etal 2016,
Chowdhury \etal 2018, Iwanek \etal 2019).

Skumanich (1972) pioneered a groundbreaking study on stellar rotation and
activity, laying the foundation for a new branch of astrophysics. One of
the key findings of Skumanich's work revolves around the analysis of
lithium abundance, indicating a decrease in its presence on stellar
surfaces as stars age, particularly during the first dredge-up
phase. Additionally, the author observed a decline in rotational velocity
with increasing stellar age, attributed to magnetic braking, where the
interaction of magnetic fields with stellar winds acts as a decelerating
force. The final finding, based on CaII emission as an indicator of
magnetic activity, revealed a diminishing trend between CaII emission
(hence magnetic activity) and stellar age. The studies initiated by
Skumanich (1972) are continued today under the name gyrochronology (Barnes
2007, Reinhold and Gizon 2015).

Measuring the rotation periods of stars is an important aspect of
astrophysics. Active regions on stellar surfaces (spots) induce periodic or
quasi-periodic brightness changes, enabling direct measurement of rotation
periods (Irwin \etal 2009). This method ranks among the most commonly used
techniques, albeit necessitating precise, long-term photometric
observations, as variability caused by spots occurs with various periods
(ranging from a fraction of a day to hundreds of days), with amplitudes
typically smaller than 0.2--0.3~mag (Lanzafame \etal 2018, Iwanek \etal
2019, Distefano \etal 2023). Another widely used method for measuring
rotation periods involves observing the CaII H and K emission lines
(Baliunas \etal 1983, Gilliland and Fisher 1985, Salabert \etal 2016).

An intriguing phenomenon is the existence of stellar magnetic activity
cycles. The first observations of the solar activity cycle were made by
Schwabe (1844, 1845) by studying the migration of sunspots on the
surface. Schwabe noticed that the sunspot migration pattern repeats
approximately every 11~yr. Studying activity cycles is difficult because we
do not directly observe spots on other stars. Additionally, such studies
require long-lasting monitoring of the same stars to cover multiple
activity cycles and measure their lengths with sufficient
precision. Despite these difficulties, such studies are being successfully
conducted (\eg Ol{\'a}h \etal 2007, 2009, Mathur \etal 2014, Ol{\'a}h \etal 2016,
Montet \etal 2017, Reinhold \etal 2017, Mart{\'i}nez \etal 2022).

Contemporary large-scale sky surveys provide invaluable data in the context
of the rotation and magnetic activity of hundreds of thousands of
stars. Particular attention should be given to the data provided by the
Gaia satellite with the Radial Velocity Spectrometer, based on which the
stellar activity index has been determined for $2\times10^6$ stars in the
Milky Way (Lanzafame \etal 2023). Reinhold \etal (2013), McQuillan \etal
(2014) and Reinhold \etal (2023) measured the rotation periods for stars
observed by the Kepler Space Telescope, while Davenport (2016) and
Davenport \etal (2019) found flaring stars in the Kepler data. The flaring
activity was also investigated using data from the Transiting Exoplanet
Survey Satellite (TESS) mission by Crowley \etal (2024).  Rotating stars
have been discovered through ground-based surveys, such as the All Sky
Automated Survey (ASAS, Kiraga 2012, Kiraga and St\k{e}pie\'n 2013), the All Sky
Automated Survey for Supernovae (ASAS-SN, Jayasinghe \etal 2020), the
Massive Compact Halo Objects (MACHO, Drake 2006) or the Optical
Gravitational Lensing Experiment (OGLE, Iwanek \etal 2019).

In this study, we use data from the OGLE project, collected for nearly
30~yr, to find as many magnetically active stars as possible. Our main goal
was to provide the astronomical community with long-term, two-band
time-series photometry, which will undoubtedly be a treasure trove of
knowledge about stellar magnetic activity. In the study by Iwanek \etal
(2019), we utilized 12\,660 objects published in this collection and
analyzed in detail the relations between the observational properties of
these stars, \ie rotation period, mean brightness, brightness amplitude,
and color index. We also demonstrated the existence of two types of spots
characterized by the dependence between changes in the color index and
changes in the brightness. This collection includes the stars analyzed in
the paper Iwanek \etal (2019) and nearly 6000 new objects classified as
rotational variables.

This paper is organized as follows. In Section~2, we describe the OGLE
project and its data. In Section~3 we outline the selection process of
rotating variables, including measuring rotation periods methods. The mean
brightness and brightness amplitude measuring methods are discussed in
Section~4. In Section~5, we describe and discuss the entire collection of
rotating variables, and we show many interesting objects with time-series
photometry collected since 1997. Conclusions can be found in Section~6.

\Section{Observations and Data Reduction}
The OGLE project can boast one of the longest-lasting large-sky survey
conducting regular high-quality photometric observations in the history of
astronomy. Historically, OGLE began its activities in 1992 (OGLE-I, Udalski
\etal 1992) using the 1.0-m Swope Telescope located at the Las Campanas
Observatory in Chile. Subsequently, with acquired funding, a dedicated
telescope for the project, the \mbox{1.3-m} Warsaw Telescope, was built at
the Las Campanas Observatory, Chile, which in 1997 initiated the second
phase of the project (OGLE-II, Udalski \etal 1997), thus beginning the
regular monitoring of stars in the Galactic bulge (BLG) and the Magellanic
Clouds. The project then went through two more phases: OGLE-III
(2001-2009, Udalski 2003) and OGLE-IV (2010-today, Udalski \etal 2015),
which mainly differed in field of view of the CCD camera used and in the
larger monitored area of the sky, including the Galactic disk. The fourth
phase of the project was interrupted for over two years in March 2020 due
to the COVID-19 pandemic. The OGLE project resumed regular observations in
August 2022.

The Warsaw Telescope is equipped with a mosaic CCD camera consisting of 32
chips, each with a resolution of ${\rm 2k\times4k}$~px, resulting in
approximately 268 million pixels. The field of view of the OGLE-IV camera
is 1.4 square degrees with a pixel scale of 0\zdot\arcs26 . As a part of
the BLG survey, we regularly observe over 400 million stars located in an
area of 182 square degrees. The OGLE-IV inner BLG footprint is presented in
Fig.~1.
\begin{figure}[htb]
\centerline{\includegraphics[width=12cm, clip=]{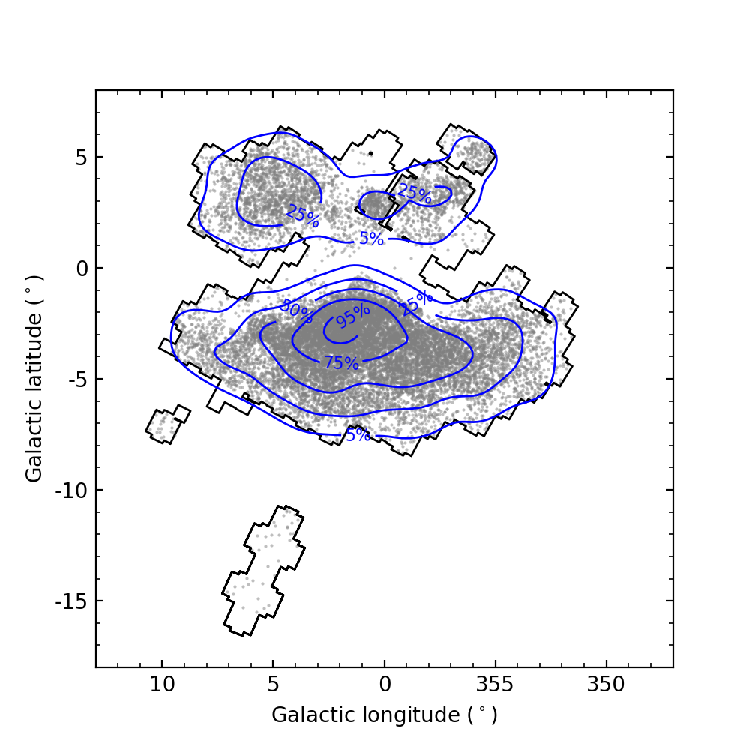}}
\FigCap{Distribution of rotating stars in the sky toward the Galactic
  bulge. Gray points denote all rotating variables discovered in the
  OGLE-IV data, while black contours correspond to the OGLE footprint. With
  blue contours, we plot the normalized density of rotating variables with
  values: 95\%, 75\%, 50\%, 25\%, and 5\%.}
\end{figure}

The OGLE observations are conducted in Johnson {\it V} (mean wavelength of
0.55~$\mu$m) and Cousins {\it I} (mean wavelength of 0.81~$\mu$m)
filters. The vast majority of observations are taken in the {\it I} band,
while {\it V}-band observations are much less frequent and generally
performed to determine the $(V-I)$ color indices of the observed stars. The
exposure times for the BLG fields are 100 s and 150 s for the {\it I}- and
{\it V}-band, respectively.

A comprehensive overview of the instrumentation, photometric reductions,
and astrometric calibrations for the OGLE-IV observations can be found in
Udalski \etal (2015).

\Section{Selection and Classification of Rotating Variables}
In large-scale sky surveys, rotating (spotted) variables are generally
treated somewhat neglectfully, and searches devoted to these stars are
rarely conducted. This is partly due to the difficulty of detecting them,
as their quasi-periodic variability usually lacks a well-defined pattern,
unlike, for example, pulsating or eclipsing stars. In the searches
conducted as a part of this work, we began with a sample of 19\,513 stars
flagged as ``spotted'' during searches for other types of variable stars
(\eg Soszy\'nski \etal 2016, Udalski \etal 2018, Wrona \etal 2022).  A
common feature in the searches for any type of variable stars in the OGLE
data is that the samples of stars are selected semi-automatically (using,
\eg some cuts on periods, amplitudes, color indices, etc.) and then they are
manually reviewed by experienced astronomers. For this reason, it was
possible to identify stars exhibiting long-term modulation of amplitude and
mean magnitude, which could potentially indicate the presence of spots on
stellar surfaces. The presence of long-term brightness modulation over
several to several dozen periods is directly related to magnetic activity
and the presence of activity cycles. Such variability can be confused, for
example, with semi-regular variations caused by radial and non-radial
pulsations in long-period variable stars (see \eg Soszy\'nski \etal 2013,
Iwanek \etal 2022).

Rotating stars are characterized by significant changes in their mean
brightness compared to their rotational variability, so any measurements of
their astrophysical properties should be conducted after removing the
long-term trend. To address this, we applied the Savitzky-Golay filter (SG,
Savitzky and Golay 1964) to remove the trend from the data. Then, we
measured the rotation periods using two methods: by utilizing the {\sf
  FNPEAKS}\footnote{\it
  http://helas.astro.uni.wroc.pl/deliverables.php?active=fnpeaks} code
based on the standard discrete Fourier transform for unevenly spaced data
(Kurtz 1985), and by employing the Fourier-like least-squares spectral
analysis proposed by Lomb (1976) and Scargle (1982), which is frequently
used in astronomy. A detailed discussion on the measurement of the rotation
period can be found in Iwanek \etal (2019).

In the first step of the selection and classification procedure, we removed
from the sample stars already classified and published in the OGLE
Collection of Variable
Stars\footnote{\it https://www.astrouw.edu.pl/ogle/ogle4/OCVS/}
(OCVS). In total, we removed 160 objects (mainly non-spotted ellipsoidal
and eclipsing variables, and several type II Cepheids). Next, we verified
each light curve (unfolded and phase-folded) through visual inspection. If
the rotation periods found using the above-described methods differed
significantly (their ratio was outside the range of 0.95-1.05), then we
phased the light curve with both periods and visually checked which one
phased the light curve better. In some cases, the rotation period required
manual fine-tuning. The selected and corrected periods were then considered
as final.

The above-mentioned procedure enhances the accuracy of the majority of
measured rotation periods through a meticulous verification process. The
visual inspection of light curves allowed us for the identification and
correction of anomalies that might eluded automated methods. By comparing
periods and fine-tuning them, we ensured that the most consistent and a
reliable value was determined. The rotation period was measured based on
the data from the fourth phase of the OGLE project, as in this phase the
light curves have the most uniform coverage. Nevertheless, we acknowledge
that the actual period may differ slightly for individual objects from the
one provided in our collection. Studying the rotation periods in each OGLE
phase separately, and even in individual observational seasons, can provide
valuable insights into changes in the rotation period and the evolution of
magnetic activity.

During this verification, we removed 910 objects from the sample because
they did not meet the criteria for rotating (spotted) variables. These were
mainly semi-regular (pulsating) variables, ellipsoidal or eclipsing (but
non-spotted) variables, artifacts from bright stars, or constant stars. We
also removed candidates for chemically peculiar stars (called Ap, or more
general CP), because they will be the subject of a separate
study. Ultimately, we classified 18\,443 stars as rotating (spotted)
variables. The in-sky distribution of all discovered rotating variables is
presented in~Fig.~1.

\vspace*{-9pt}
\Section{Determination of Mean Magnitudes and Amplitudes}
\vspace*{-5pt}

For the classified stars, we measured the mean magnitudes in both the {\it
  I}- and {\it V}-bands (if the light curve in the {\it V}-band was
available). First, we transformed the light curve from magnitudes to
flux. Then, we removed the trend using the SG filter, and we rejected
outliers deviating by more than 3$\sigma$ from the mean flux. After these
preparation steps, we measured the mean flux and converted back the light curves
and mean flux to magnitudes.

To measure the brightness amplitude in both OGLE bands (if the light curve
in the {\it V}-band was available), we used the method described in Iwanek
\etal (2019). We divided the light curve into bins containing at least 200
epochs (if there were more than 200 epochs, otherwise all epochs
constituted one bin) and covering at least five rotation cycles, as Mathur
\etal (2014) identified such bins as optimal for active stars. Then, we
calculated the difference between the 95th and 5th percentiles of the
brightness distribution in each bin. The final amplitude was determined as
the median of differences calculated in all bins, divided by the distance
between the above-mentioned percentiles, which is 0.9. Both parameters, \ie
the mean magnitude and amplitudes were measured based on the OGLE-IV light
curves, as in this phase the light curves have the most uniform coverage.

\vspace*{-9pt}
\Section{The OGLE Collection of Rotating Variables toward the Galactic Bulge}
\vspace*{-5pt}

We present the OGLE collection of rotating (spotted) stars discovered
toward the Galactic bulge, which ultimately contains 18\,443 objects. This
is the first cata-

\newpage
\begin{landscape}
\MakeTableSep{lccllll}{12.5cm}{File {\tt ident.dat} with equatorial 
coordinates and identifications for all rotating variables}
{
\hline
\noalign{\vskip3pt}
ID & R.A. & Decl. & OGLE-IV ID & OGLE-III ID & OGLE-II ID & Other ID \\
 & [h:m:s] & [deg:m:s] &  &  &  &  \\ 
\noalign{\vskip3pt}
\hline
\noalign{\vskip3pt}
OGLE-BLG-ROT-000001 & 17:12:53.64 & $-$29:41:24.2 & BLG617.25.83018  & BLG363.7.41109 & \\      
OGLE-BLG-ROT-000002 & 17:13:00.53 & $-$29:28:55.8 & BLG617.25.76720  & BLG363.6.82514 & \\ 
OGLE-BLG-ROT-000003 & 17:13:01.76 & $-$29:52:00.3 & BLG617.16.68144  & BLG363.8.12132 & \\ 
OGLE-BLG-ROT-000004 & 17:13:03.40 & $-$29:34:19.8 & BLG617.25.67272  & BLG363.6.13132 & \\ 
OGLE-BLG-ROT-000005 & 17:13:03.60 & $-$28:13:57.7 & BLG616.25.77048  & &  \\ 
OGLE-BLG-ROT-000006 & 17:13:08.32 & $-$29:46:43.2 & BLG617.16.76016  & BLG363.8.83264 & \\ 
OGLE-BLG-ROT-000007 & 17:13:21.10 & $-$29:54:06.0 & BLG617.16.9711   & &  \\ 
OGLE-BLG-ROT-000008 & 17:13:25.16 & $-$29:29:53.6 & BLG617.25.17094  & BLG363.6.89829 & \\ 
OGLE-BLG-ROT-000009 & 17:13:34.90 & $-$29:34:42.6 & BLG617.24.101118 & BLG363.6.26424 & \\      
OGLE-BLG-ROT-000010 & 17:13:35.73 & $-$28:24:47.9 & BLG616.24.67136  & & \\              
OGLE-BLG-ROT-000011 & 17:13:38.92 & $-$29:45:09.2 & BLG617.15.107872 & BLG363.8.125665 & \\
OGLE-BLG-ROT-000012 & 17:13:40.17 & $-$29:18:14.4 & BLG617.32.85672  & BLG363.5.129784 & \\
OGLE-BLG-ROT-000013 & 17:13:41.96 & $-$30:15:53.3 & BLG617.07.60080  & & \\
OGLE-BLG-ROT-000014 & 17:13:42.34 & $-$29:48:57.9 & BLG617.15.74001  & BLG363.8.62592 & \\
OGLE-BLG-ROT-000015 & 17:13:42.80 & $-$28:00:37.6 & BLG616.32.66570  & & \\
OGLE-BLG-ROT-000016 & 17:13:45.29 & $-$29:34:05.5 & BLG617.24.75365  & BLG363.6.30704 & \\
OGLE-BLG-ROT-000017 & 17:13:51.20 & $-$28:12:23.4 & BLG616.24.92498  & & \\
OGLE-BLG-ROT-000018 & 17:13:51.46 & $-$29:22:46.3 & BLG617.32.53344  & BLG363.4.40825 & \\
OGLE-BLG-ROT-000019 & 17:13:52.40 & $-$29:43:22.3 & BLG617.24.30037  & BLG363.2.1704  & \\
OGLE-BLG-ROT-000020 & 17:13:52.50 & $-$29:53:54.4 & BLG617.15.37841  & & \\
\noalign{\vskip3pt}
\hline
\noalign{\vskip7pt}
\multicolumn{5}{p{12cm}}{Table presents only the first twenty rows for
  guidance regarding their form and content.}}
\end{landscape}

\noindent
log of rotating variables in the history of astronomy with
such long (at least 12~yr) time-series photometry in {\it I}- and {\it
  V}-band. It is indisputably one of the best existing photometric datasets
for studying stellar activity, including activity cycles. The data are
available through the OGLE website:

\begin{itemize}
\item {\it https://www.astrouw.edu.pl/ogle/ogle4/OCVS/blg/rot/} -- the whole dataset.
\item {\it https://ogledb.astrouw.edu.pl/ogle/OCVS/} -- user-friendly search engine for the entire OGLE Collection of Variable Stars.
\end{itemize}

\MakeTable{lccccc}{10cm}{File {\tt rot.dat} containing mean magnitudes and brightness
  amplitudes in {\it V}- and {\it I}-bands, as well as rotation periods}
{\hline
\noalign{\vskip3pt}
ID & $V$ & $A_V$ & $I$ & $A_I$ & $P$ \\
 & [mag] & [mag] & [mag] & [mag] & [d] \\ 
\noalign{\vskip3pt}
\hline
\noalign{\vskip3pt}
OGLE-BLG-ROT-000001 &  17.989 &  0.208 &  16.026 &  0.167 & 38.760531 \\
OGLE-BLG-ROT-000002 &  20.947 &  0.786 &  19.163 &  0.646 &  3.927652 \\
OGLE-BLG-ROT-000003 &  17.929 &  0.556 &  15.928 &  0.192 & 86.192261 \\
OGLE-BLG-ROT-000004 &  18.054 &  0.187 &  16.341 &  0.126 & 35.143542 \\
OGLE-BLG-ROT-000005 &  17.440 &  0.222 &  15.603 &  0.203 & 51.283989 \\
OGLE-BLG-ROT-000006 &  19.852 &  0.206 &  18.013 &  0.191 &  4.294912 \\
OGLE-BLG-ROT-000007 &  19.646 &  0.478 &  17.571 &  0.374 & 13.898817 \\
OGLE-BLG-ROT-000008 &  19.565 &  0.308 &  17.978 &  0.211 &  7.720678 \\
OGLE-BLG-ROT-000009 &  19.943 &  0.383 &  17.974 &  0.267 &  2.094063 \\
OGLE-BLG-ROT-000010 &  21.110 &  1.160 &  18.953 &  0.576 &  4.101264 \\
OGLE-BLG-ROT-000011 &  18.659 &  0.519 &  16.713 &  0.159 & 29.043105 \\
OGLE-BLG-ROT-000012 &  19.571 &  0.349 &  17.609 &  0.299 & 11.729599 \\
OGLE-BLG-ROT-000013 &  17.997 &  0.215 &  15.998 &  0.163 & 86.192261 \\
OGLE-BLG-ROT-000014 &  19.107 &  0.307 &  17.178 &  0.188 & 41.002912 \\
OGLE-BLG-ROT-000015 &  18.886 &  0.388 &  16.909 &  0.327 & 21.418297 \\
OGLE-BLG-ROT-000016 &  18.485 &  0.253 &  16.709 &  0.162 & 70.625070 \\
OGLE-BLG-ROT-000017 &  19.813 &  0.304 &  18.206 &  0.382 &  4.825298 \\
OGLE-BLG-ROT-000018 &  19.303 &  0.332 &  17.501 &  0.261 & 10.284098 \\
OGLE-BLG-ROT-000019 &  --     &  --    &  18.895 &  0.516 &  2.565629 \\
OGLE-BLG-ROT-000020 &  18.498 &  0.142  & 16.626 &  0.159 & 18.791846 \\
\noalign{\vskip3pt}
\hline
\noalign{\vskip3pt}
\multicolumn{6}{p{10cm}}{The sign "$-$" in the columns of $V$ and $A_V$ 
means, that the light curve in the {\it V}-band is not available. 
Table presents only the first twenty rows for guidance regarding their form and content.}
}
The websites of the OGLE collection of rotating variables are structured as
follows. The file {\tt ident.dat} contains the list of stars with their
unique identifier: OGLE-BLG-ROT-NNNNNN, where NNNNNN is a six-digit number,
J2000 equatorial coordinates, identifications in the OGLE-II, OGLE-III and
OGLE-IV databases, and designations taken from external catalogs, \ie the
International Variable Star Index\footnote{\it
  https://www.aavso.org/vsx/index.php} (VSX, Watson \etal 2006) and ASAS-SN
(Shappee \etal 2014, Jayasinghe \etal 2020).  In the file {\tt rot.dat},
we provide observational parameters for each star -- the rotation period,
mean magnitudes and amplitudes in the {\it I}- and {\it V}-band. The file
{\tt remarks.txt} contains comments on interesting stars (with identifiers
from OCVS). The file {\tt remarks.txt} will be updated in the future on
some interesting, unusual objects. In Tables 1, 2 and 3, we present the
first twenty rows from the files {\tt ident.dat}, {\tt rot.dat}, and {\tt
  remarks.txt}, respectively, for guidance regarding their form and
content.

\MakeTable{lcc}{10cm}{File {\tt remarks.txt} containing comments on 
some interesting objects, with identifiers from other OGLE catalogs}
{
\hline
\noalign{\vskip3pt}
ID & remark & other OGLE ID  \\ 
\noalign{\vskip3pt}
\hline
\noalign{\vskip3pt}
OGLE-BLG-ROT-000965 & RSCVn  & OGLE-BLG-ECL-032766 \\
OGLE-BLG-ROT-001092 & RSCVn  & OGLE-BLG-ECL-036149 \\
OGLE-BLG-ROT-001820 & RSCVn  & OGLE-BLG-ECL-055958 \\
OGLE-BLG-ROT-003288 & RSCVn  & OGLE-BLG-ECL-095740 \\
OGLE-BLG-ROT-004213 & flares & \\
OGLE-BLG-ROT-004667 & flares & \\
OGLE-BLG-ROT-004832 & flares & \\
OGLE-BLG-ROT-005253 & flares & \\
OGLE-BLG-ROT-005282 & RSCVn  & \\
OGLE-BLG-ROT-005373 & flares & \\
OGLE-BLG-ROT-005511 & flares & \\
OGLE-BLG-ROT-005552 & flares & \\
OGLE-BLG-ROT-005594 & flares & \\
OGLE-BLG-ROT-005784 & flares & \\
OGLE-BLG-ROT-005997 & flares & \\
OGLE-BLG-ROT-006123 & RSCVn  & \\
OGLE-BLG-ROT-006165 & flares & \\
OGLE-BLG-ROT-006255 & flares & \\
OGLE-BLG-ROT-006327 & flares & \\
OGLE-BLG-ROT-006329 & RSCVn  & OGLE BLG-ECL-170709  \\
\noalign{\vskip3pt}
\hline
\multicolumn{3}{p{8cm}}{Table presents only the first twenty rows for guidance
  regarding their form and content.}
}

\MakeTable{lcccc}{9cm}{Statistics about the number of epochs in the light curves,
    separately for each OGLE phase, with division into bands}
{
\hline
\noalign{\vskip3pt}
Phase & Band  & Med. & Max. & No. of light curves  \\ 
\noalign{\vskip3pt}
\hline
\noalign{\vskip3pt}
OGLE-II  & {\it I} & 263  & 537     & 1550 \\
OGLE-III & {\it I} & 629  & 2540    & 13\,532 \\
OGLE-IV  & {\it I} & 2053 & 17\,682 & 18\,443 \\ 
\noalign{\vskip3pt}
\hline
\noalign{\vskip3pt}
OGLE-II  & {\it V} &    8 & 18      & 1476 \\
OGLE-III & {\it V} &    5 & 41      & 13\,119 \\
OGLE-IV  & {\it V} &  144 & 253     & 17\,873 \\
\noalign{\vskip3pt}
\hline
}

\begin{figure}[htb]
\vglue-11pt
\includegraphics[width=13cm]{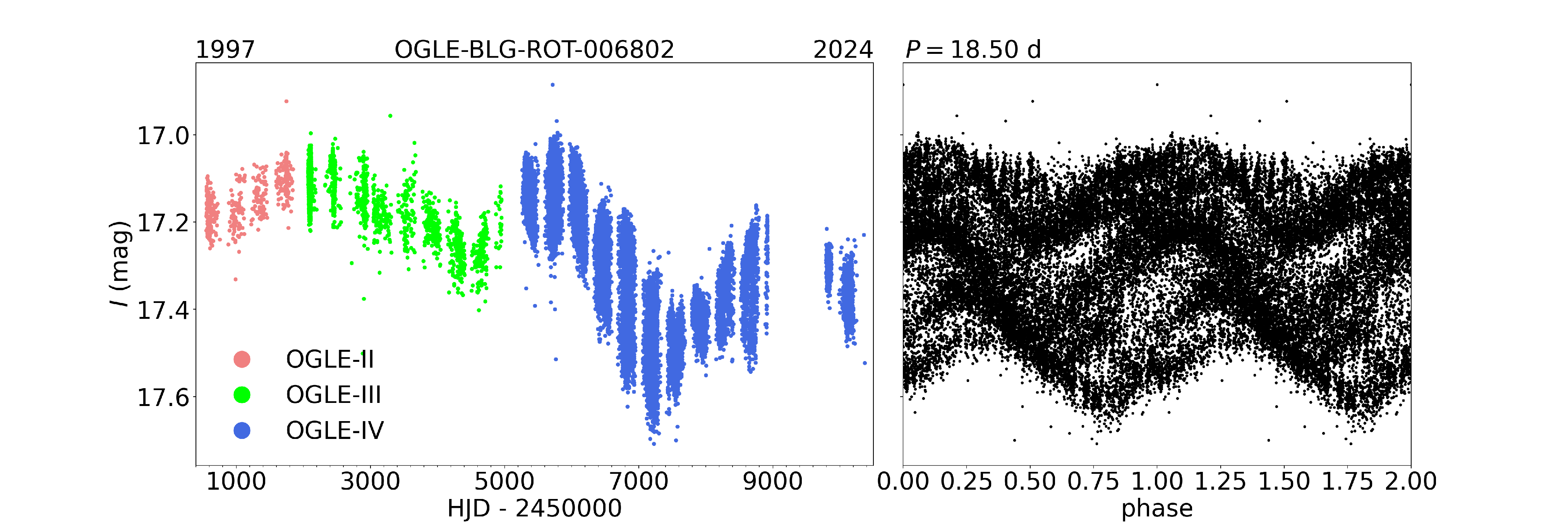}
\includegraphics[width=13cm]{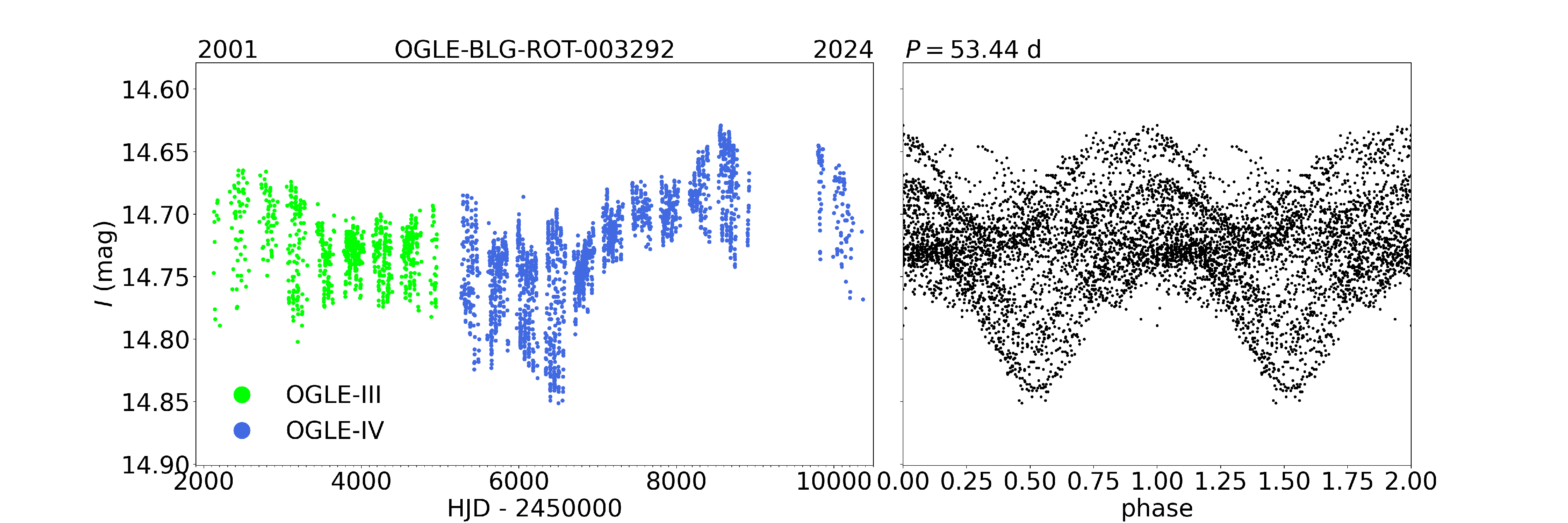}
\includegraphics[width=13cm]{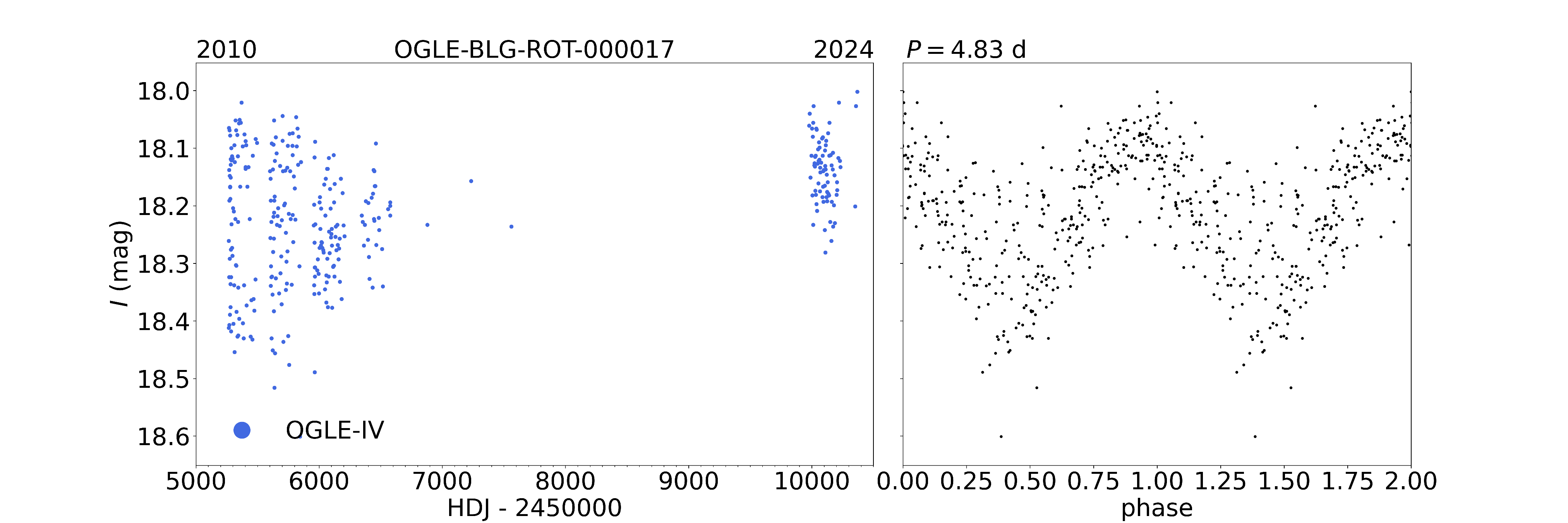}
\FigCap{Three examples of rotating variables from our collection, ordered
  by the number of epochs in light curves. {\it Top row:} a well-covered light
  curve of rotating variable OGLE-BLG-ROT-006802 containing 19\,939 epochs
  that span 27~yr (from 1997 to 2024). {\it Middle row:} a medium-covered
  light curve of rotating variables OGLE-BLG-ROT-003292 containing 3173
  epochs that span 23~yr (from 2001 to 2024). {\it Bottom row:} a
  poorly-covered light curve of rotating variable OGLE-BLG-ROT-000017
  containing 333 epochs that span 14~yr (from 2010 to 2024). In the
  {\it left panels}, we show unfolded light curves, while in the
  {\it right panels}, we show phase-folded light curves with rotation
  period $P$ indicated above the plots. The dates at the tops of the
  unfolded light curves mark the year when the observations started and the
  year of the last used observations in the collection. Different OGLE
  phases are marked with different colors (red, green, and blue).}
\end{figure}

The subdirectory {\tt fcharts/} contains finding charts for all
objects. Each finding chart is a $60\arcm\times60\arcm$ subframe of the
OGLE {\it I}-band reference image, oriented with north at the top and east
at the left. The object is identified with a white cross in the center of
the finding chart.

In the subdirectories {\tt phot\_ogle2/}, {\tt phot\_ogle3/}, and
{\tt phot\_ogle4/}, we provide all available time-series photometry in
the {\it I}-band and {\it V}-band (if available) from the OGLE-II
(1997-2000), OGLE-III (2001-2009), and OGLE-IV (2010-2024) phases,
respectively. The data from different phases were calibrated separately to
the standard Johnson-Cousins photometric system. When using the OGLE data,
it is necessary to take into account that there may be a zero-point offset
between the light curves from different phases. Usually, the uncertainties
of the OGLE photometric calibrations do not exceed 0.05~mag. If it is
necessary to merge the light curves, such an offset should be considered in
the analysis.

During the preparation of the data for the catalog, we decided not to
remove any outlying measurements from the light curves published in this
collection. In the case of rotating stars, this is particularly challenging
as flares or eclipses may occur, which can appear as outliers but may not
be so. Therefore, during analyses using data from the collection, we
encourage the reader to independently assess whether outlying measurements
are genuine phenomena or just outliers, that could be removed.

The {\it I}-band light curves are available for all variables from the
collection, while the {\it V}-band are available for 97\% rotating stars
(17\,873 objects). Statistics regarding the light curves in both bands from
the various phases of the OGLE project are presented in Table~4.  In
Fig.~2, we present three example light curves of rotating variables, with
different coverage of the OGLE observations.

\subsection{Crossmatch with Other Catalogs of Variable Stars}
We crossmatched our sample of rotating variables with two external
catalogs: VSX and ASAS-SN. We searched for objects around our stars'
coordinates within a radius of 5\arcs\ for VSX and 1\arcs\ for ASAS-SN.

We identified 350 counterparts to our objects in the VSX. Most of these
objects (342 out of 350) are classified in the VSX as ROT (rotating
variables) or RS (RS~CVn-type variables). We checked stars with different
classifications and confirmed that all of them are rotating variables. For
316 stars, the periods reported in the VSX are consistent with the periods
measured in this paper to within 5\%. For the remaining stars, we checked
both periods and confirmed that the ones we measured were correct.

In the ASAS-SN catalog, we found 175 counterparts to objects from our
collection. Most of the objects (166 out of 175) are classified by the
ASAS-SN team as ROT (rotating variables) or VAR (variable stars). We
checked the remaining nine objects and confirmed that all of them are
rotating variables. Only 37 stars have a period reported in the ASAS-SN
catalog, and 8 of them have a period consistent with the period measured by
us. We checked the remaining 29 objects and confirmed that the periods we
measured were correct.

\subsection{Observational Parameters of Rotating Stars}
In Fig.~3, we present the distributions of rotation periods (left panel),
mean brightnesses (middle panel), and amplitudes (right panel). Stars with
rotation periods shorter than $\approx2$~d typically have smaller amplitudes
(up to about 0.2~mag) and are mostly dwarfs. Giants begin to dominate from
a rotation period longer than 2~d onwards (see \eg Fig.~11 in Iwanek \etal
2019). The peak of the rotation period distribution is around 50~d. 

The wide range of rotation periods suggests that the level of magnetic
activity is not dependent on the rotation velocity. The truncation of the
distribution at a rotation period of just over 100~d is a result of the
selection effect and arises from the fact that this collection of spotted
variables is mostly a by-product of searches for variables of other types.
\begin{figure}[h]
\includegraphics[width=12.5cm]{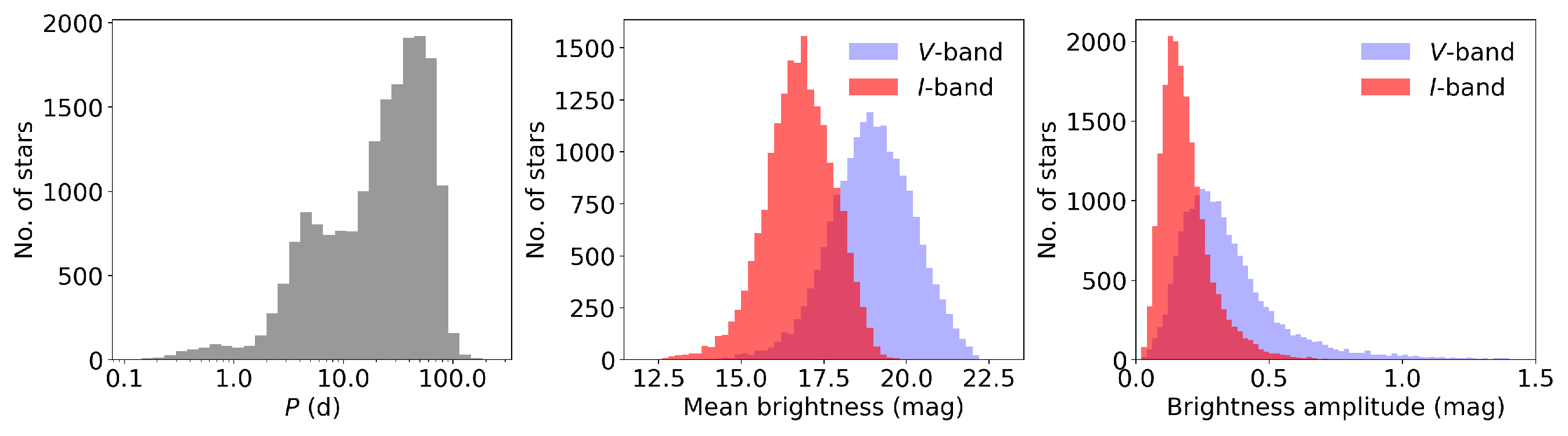}
\FigCap{Distributions of observational parameters of rotating
  variables. {\it Left panel:} distribution of rotation period. {\it Middle panel:}
  distribution of mean brightness in the {\it I}-band and {\it
    V}-band. {\it Right panel:} distribution of amplitude in the {\it I}-band and
  {\it V}-band.}
\end{figure}

The mean brightness range in the {\it I}-band is between $\approx12.5$~mag
and 20~mag, while in the {\it V}-band the mean brightness ranging between
$\approx15$~mag and 22.5~mag. The peak of mean brightness distribution in
the {\it I}-band is around 16.9 mag, while in the {\it V}-band, it is at
18.9~mag.

The distribution of amplitudes is much narrower in the {\it I}-band than in
the \mbox{{\it V}-band}. In the {\it I}-band, we observe amplitudes ranging from a
few mmag to about 0.7~mag, with a peak at 0.13~mag. For the {\it V}-band,
the distribution is much broader, ranging from a few mmag to over 1~mag,
with a peak at 0.25~mag. In the \mbox{{\it V}-band}, active regions (mainly spots)
have a greater impact on the total light reaching us from the star. The
width of the amplitude distribution in the {\it V}-band relative to the
{\it I}-band may be slightly distorted due to the significantly smaller
number of measurements in the {\it V}-band, and thus less well-measured
amplitudes.

\subsection{Purity and Completeness of the Collection}
The purity of our collection is very high due to the method of selecting
rotating stars, \ie visual inspection and final confirmation of
classification correctness by the human eye. As for completeness, we do not
even attempt to estimate it. Recent studies show that even 8\% of giants
exhibit magnetic activity (Gaulme \etal 2020). Additionally, magnetic
activity is a dynamic process with a relatively short time scale compared
to other astrophysical processes, \eg pulsations. It can fade away and
reappear over time, so detecting it depends on the time when we observe a
given star. An example of such a phenomenon is the well-known Maunder
Minimum, which occurred from 1645 to 1715, during which a significantly
lower number of sunspots (by several orders of magnitude) were observed
compared to a similar period in modern times (Sp\"orer 1889, Maunder 1922,
Eddy 1976, Hayakawa \etal 2024). The completeness of this collection can
certainly be improved, and we intend to supplement the collection in the
future.

\subsection{Stars with over 20 years of Photometric Data}
In our collection of rotating stars, 1512 objects have been continuously
observed since 1997. This means that for each of these stars, time-series
photometry is available from the second, third, and fourth phases of the
OGLE project. For 12\,020 objects we have data from OGLE-III and OGLE-IV
phases, covering the time range from 2001 to 2024. For nearly one-third of
the variables in the collection, the data are available only from the
fourth phase of the OGLE project, which spans over 12~yr of continuous
observations (from 2010 to 2024).

Almost all the light curves published in this collection are ideal datasets
for measuring and studying activity cycles. Reinhold \etal (2017) studying
stars from the Kepler satellite, detected activity cycles with lengths
ranging from 0.5~yr to 6~yr. The time span of the OGLE observations and
their dense coverage of the light curves make this dataset unique and may
allow the detection of activity cycles even as long as the solar activity
cycle, \ie 11~yr or more. In Figs.~4--10, we present 35 examples of
rotating variables with the longest time-series photometry, \ie since 1997
to 2024. 

In Figs. 4--10, we present 35 examples of rotating variables with the
longest time-series photometry, \ie since 1997 to 2024. In the left panels,
we show unfolded light curves, while in the right panels, we show
phase-folded light curves with rotation period $P$ indicated above the
plots. The dates at the tops of the unfolded light curves mark the year
when the observations started and the year of the last used observations in
the collection. Different OGLE phases are marked with different colors
(red, green, and blue).

\begin{figure}[p]
\centerline{\includegraphics[width=11.6cm]{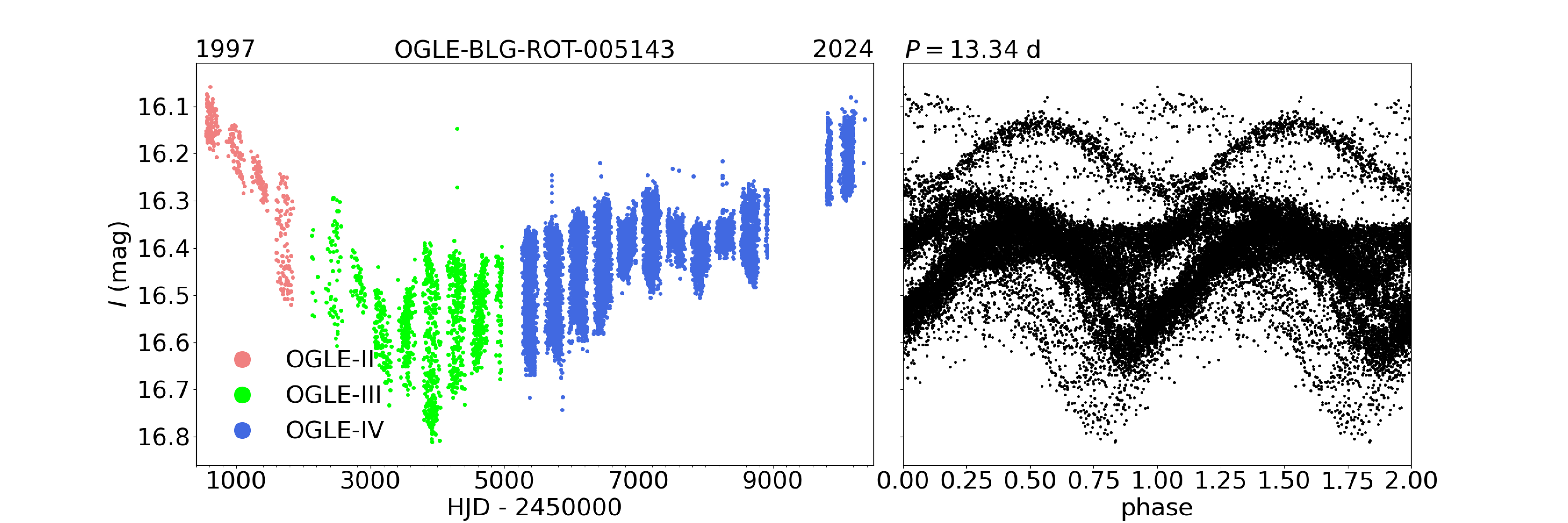}}
\centerline{\includegraphics[width=11.6cm]{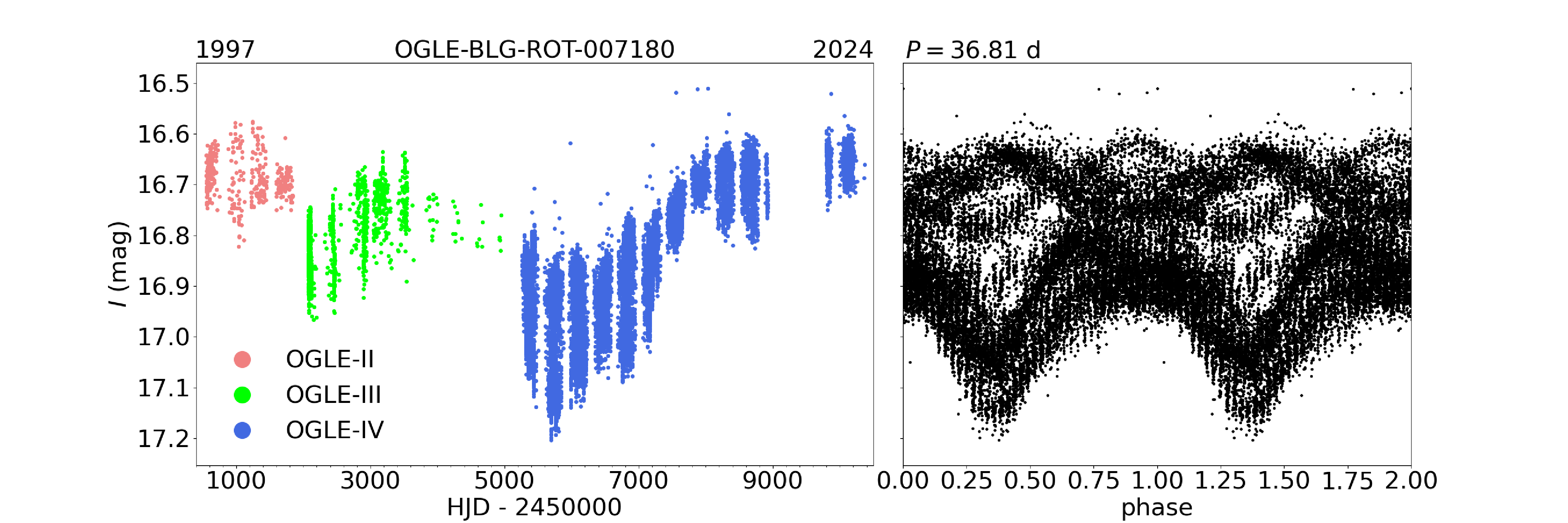}}
\centerline{\includegraphics[width=11.6cm]{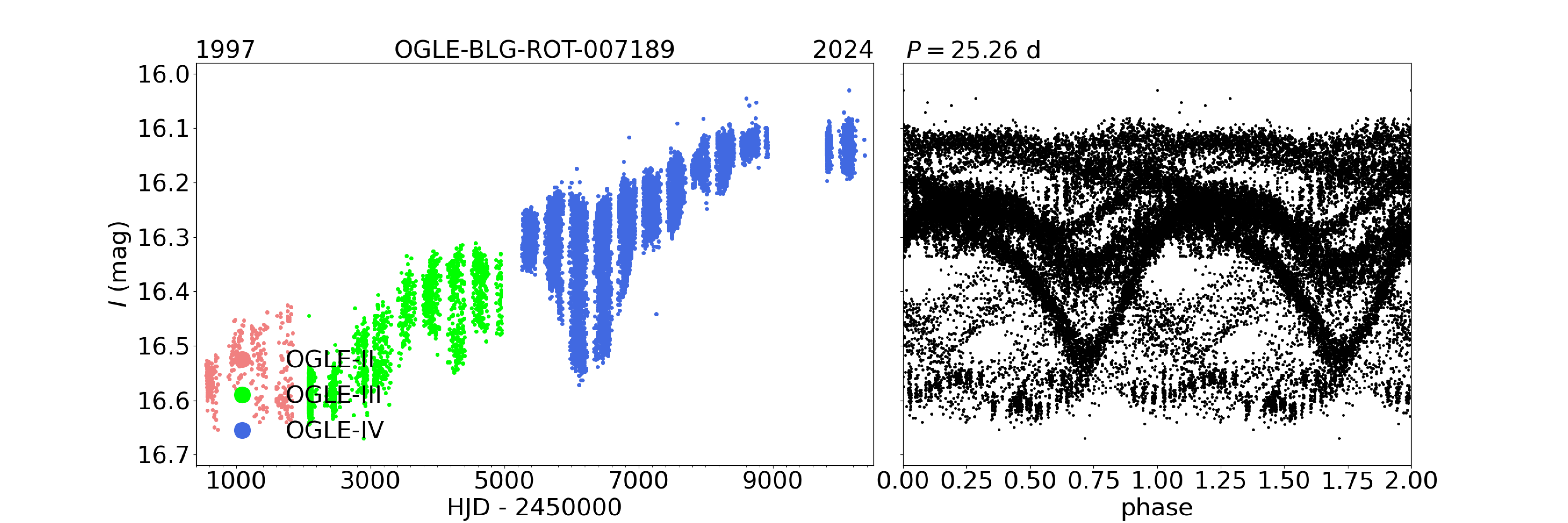}}
\centerline{\includegraphics[width=11.6cm]{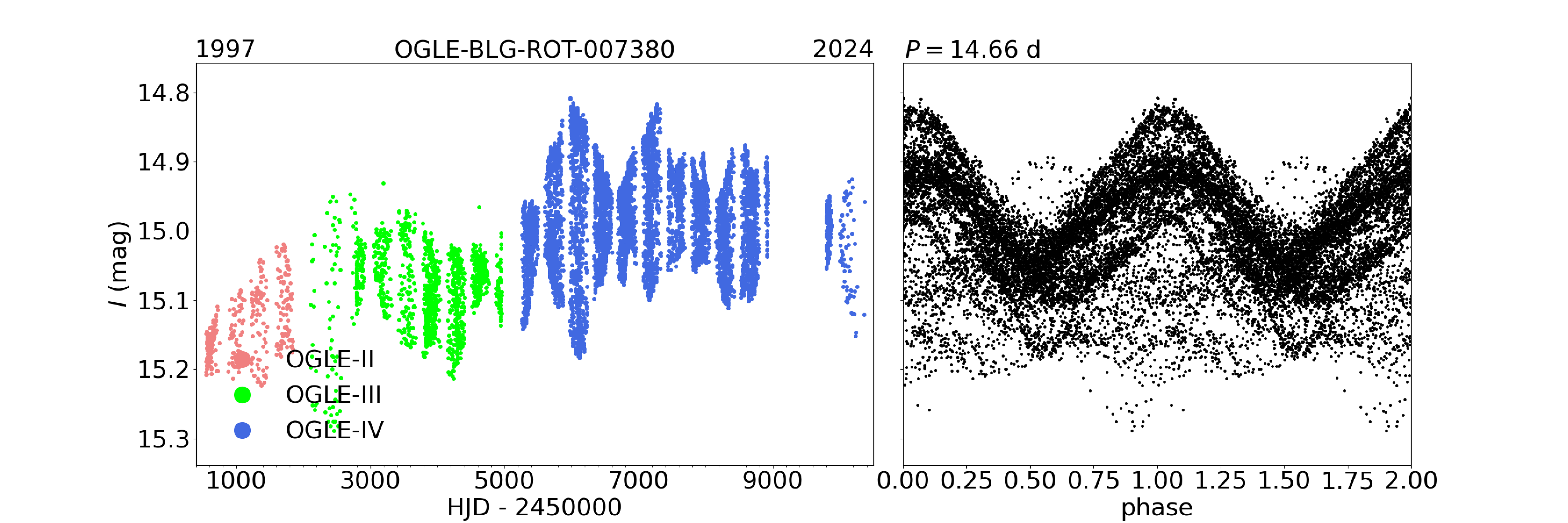}}
\centerline{\includegraphics[width=11.6cm]{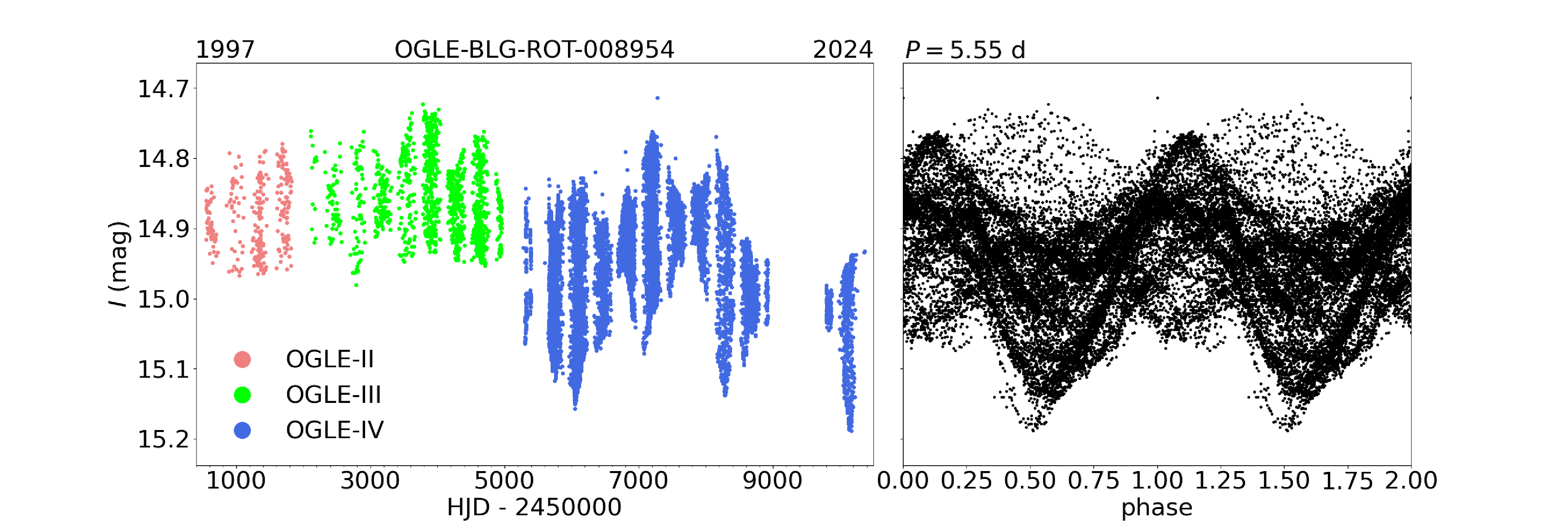}}
\vskip3pt
\FigCap{Five examples of rotating variables from our collection, with the
  longest time-series photometry, spanning from 1997 to 2024.}
\end{figure}

\begin{figure}[p]
\centerline{\includegraphics[width=11.6cm]{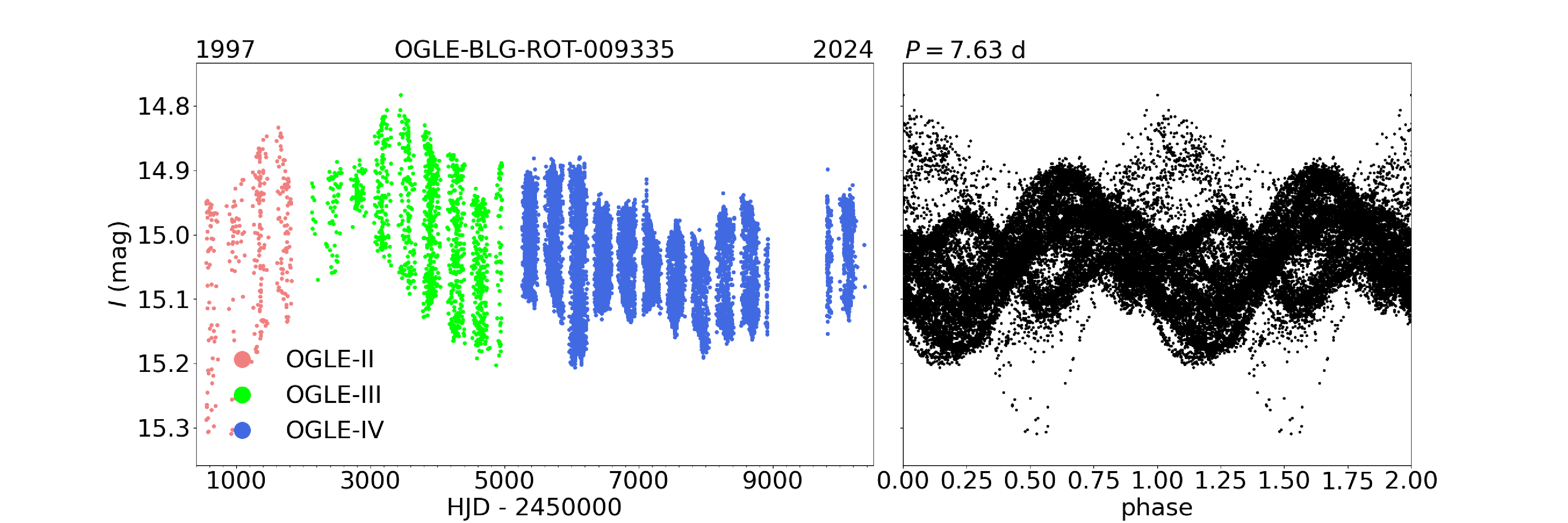}}
\centerline{\includegraphics[width=11.6cm]{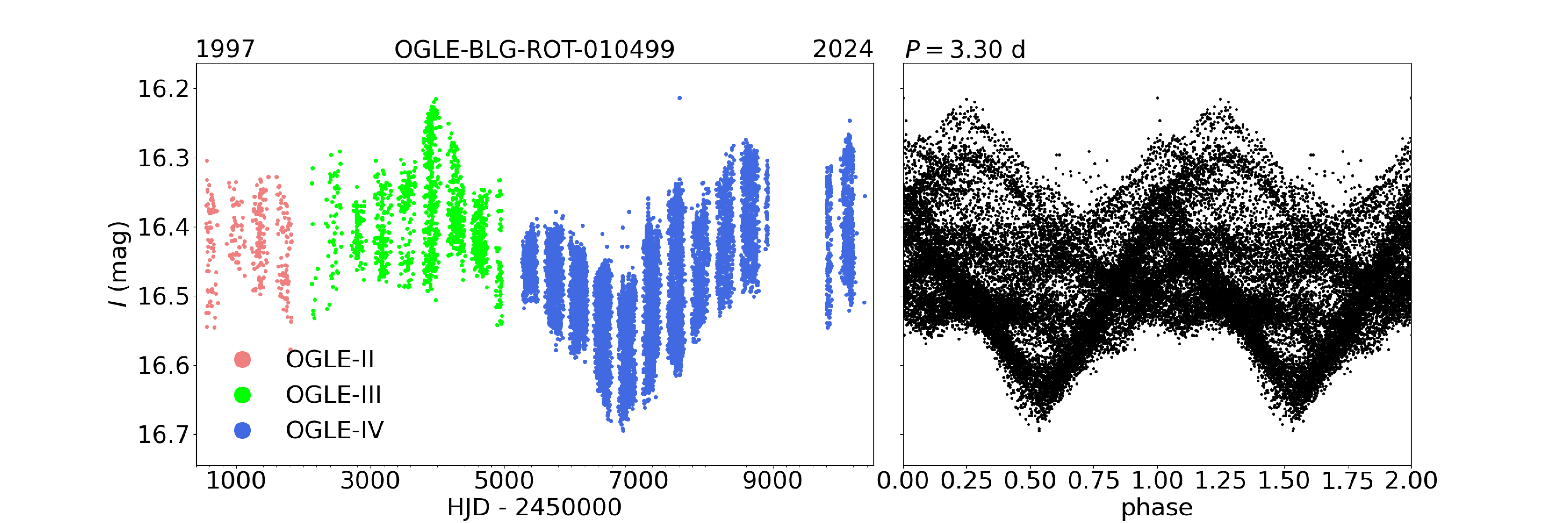}}
\centerline{\includegraphics[width=11.6cm]{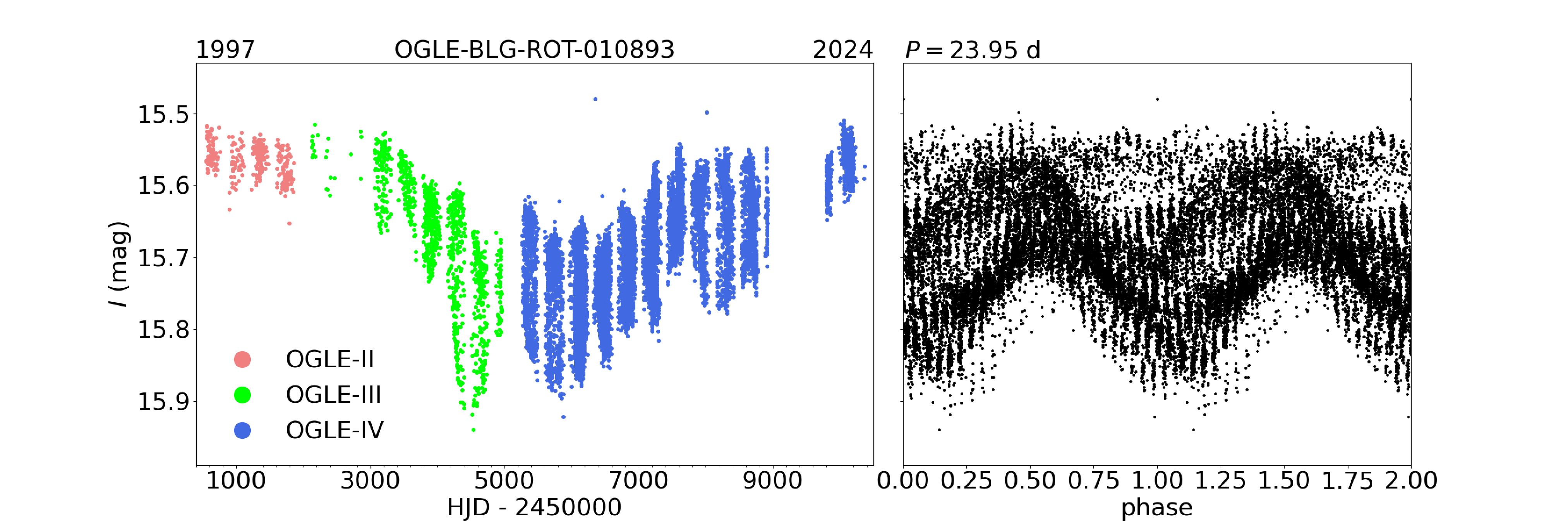}}
\centerline{\includegraphics[width=11.6cm]{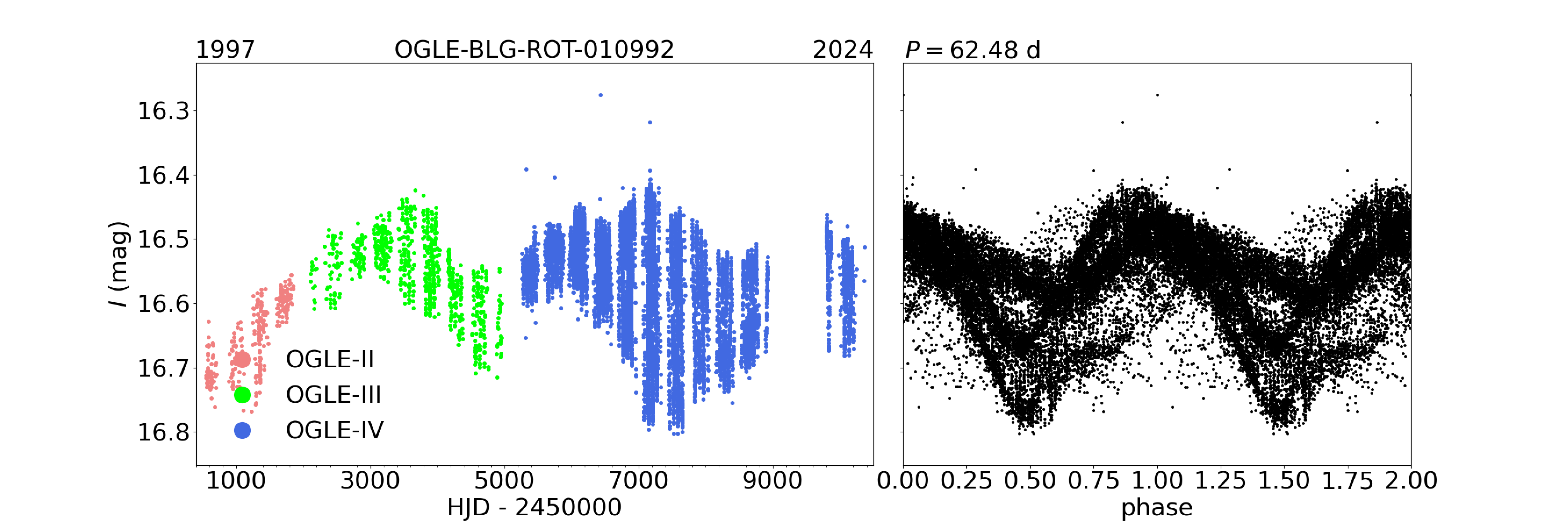}}
\centerline{\includegraphics[width=11.6cm]{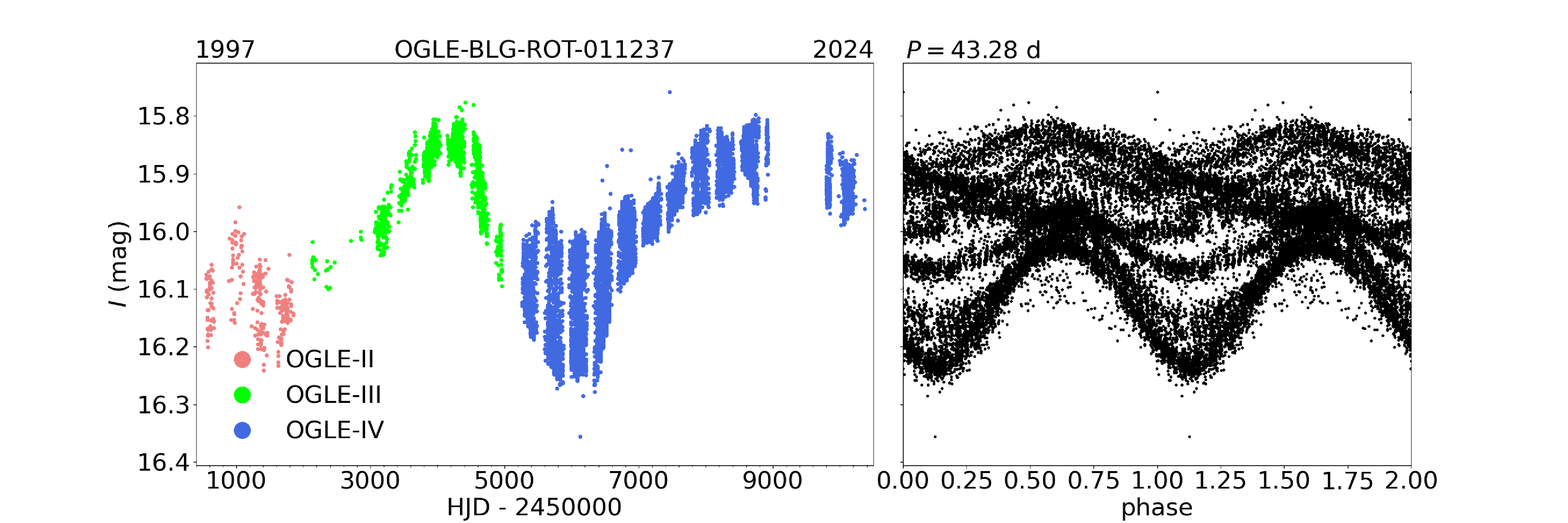}}
\vskip3pt\FigCap{Same as Fig.~4, but five other examples of rotating variables are presented.}
\end{figure}

\begin{figure}[p]
\centerline{\includegraphics[width=11.6cm]{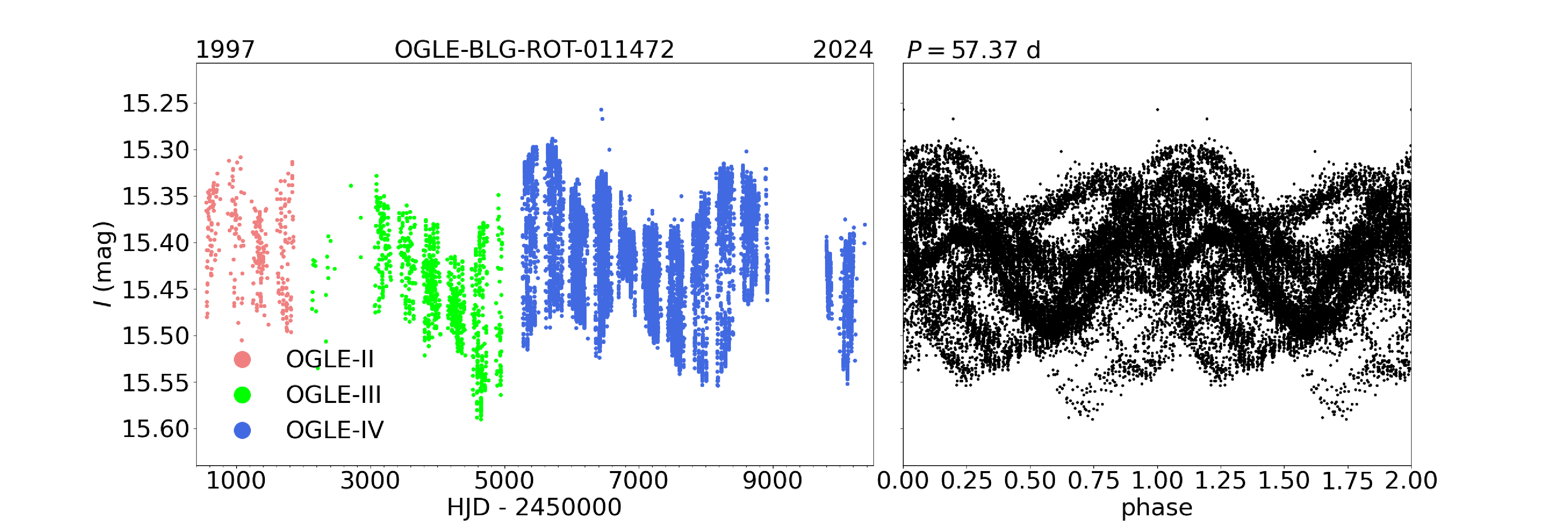}}
\centerline{\includegraphics[width=11.6cm]{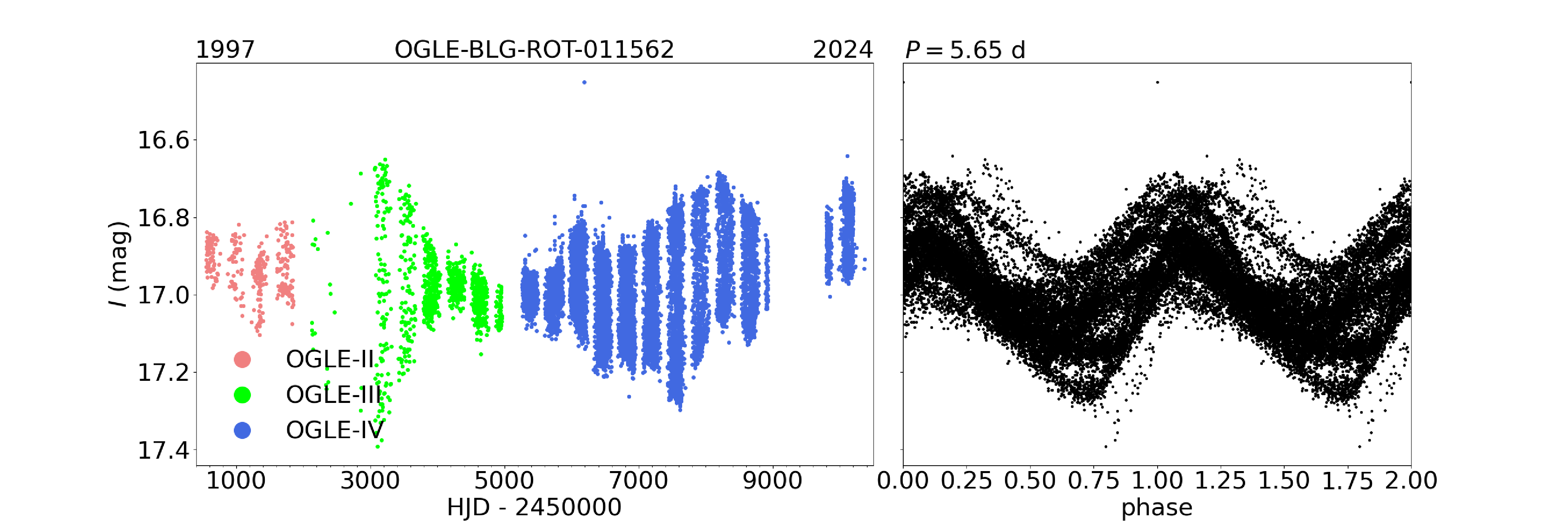}}
\centerline{\includegraphics[width=11.6cm]{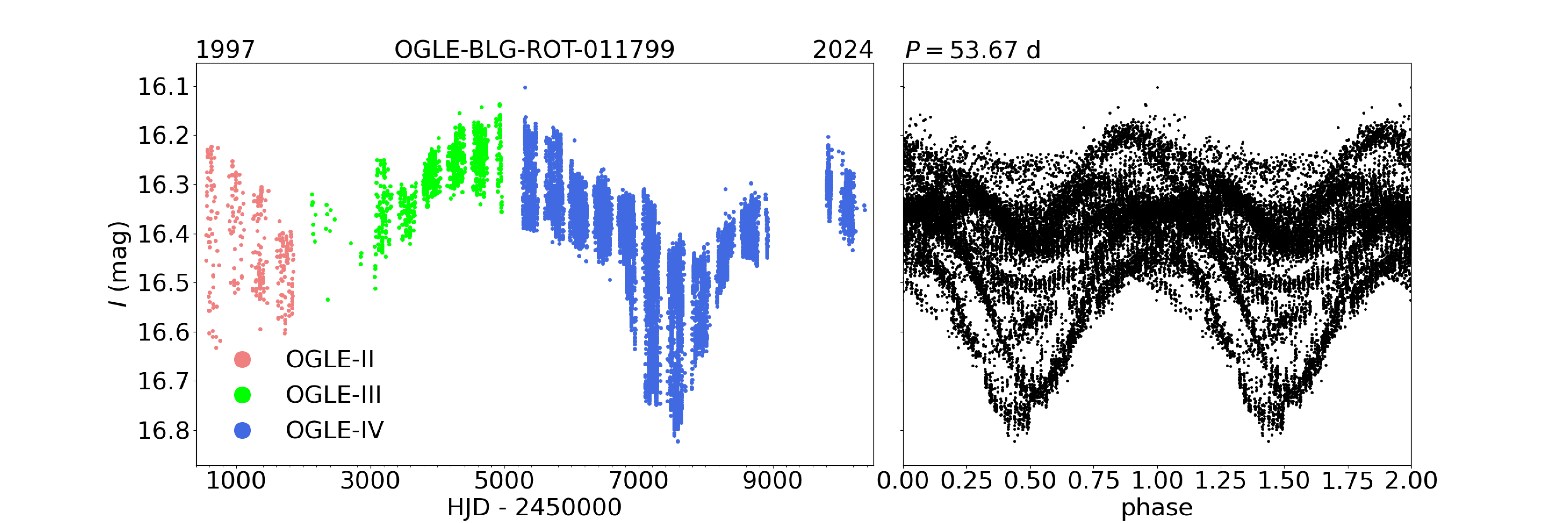}}
\centerline{\includegraphics[width=11.6cm]{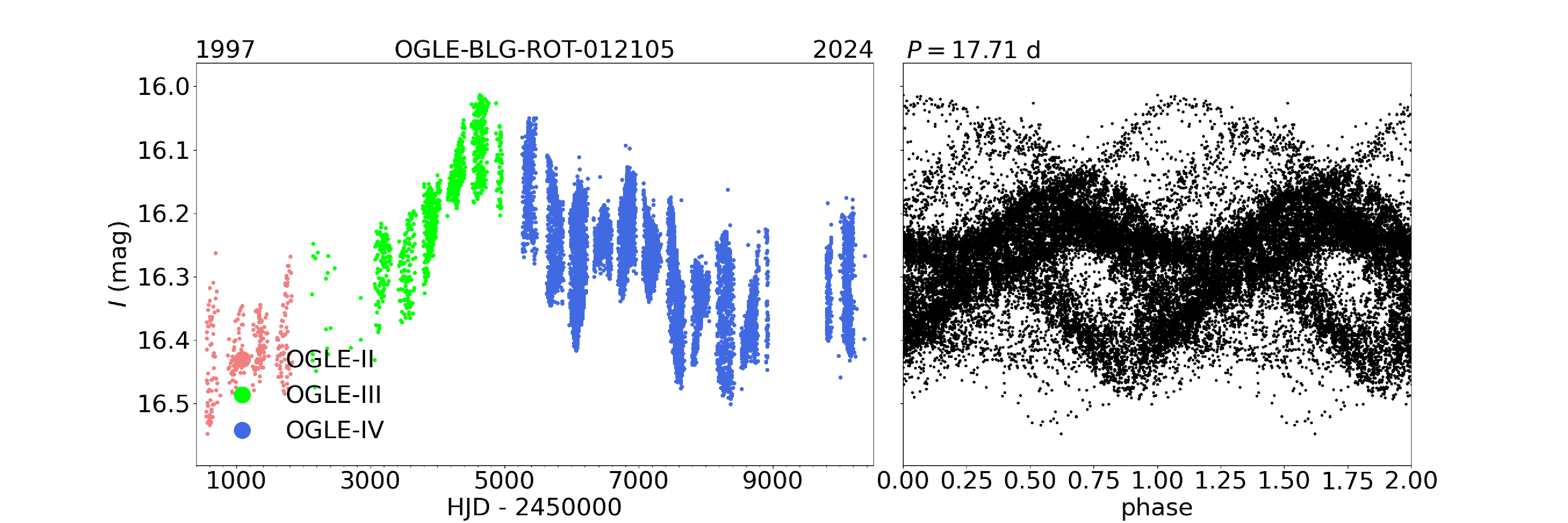}}
\centerline{\includegraphics[width=11.6cm]{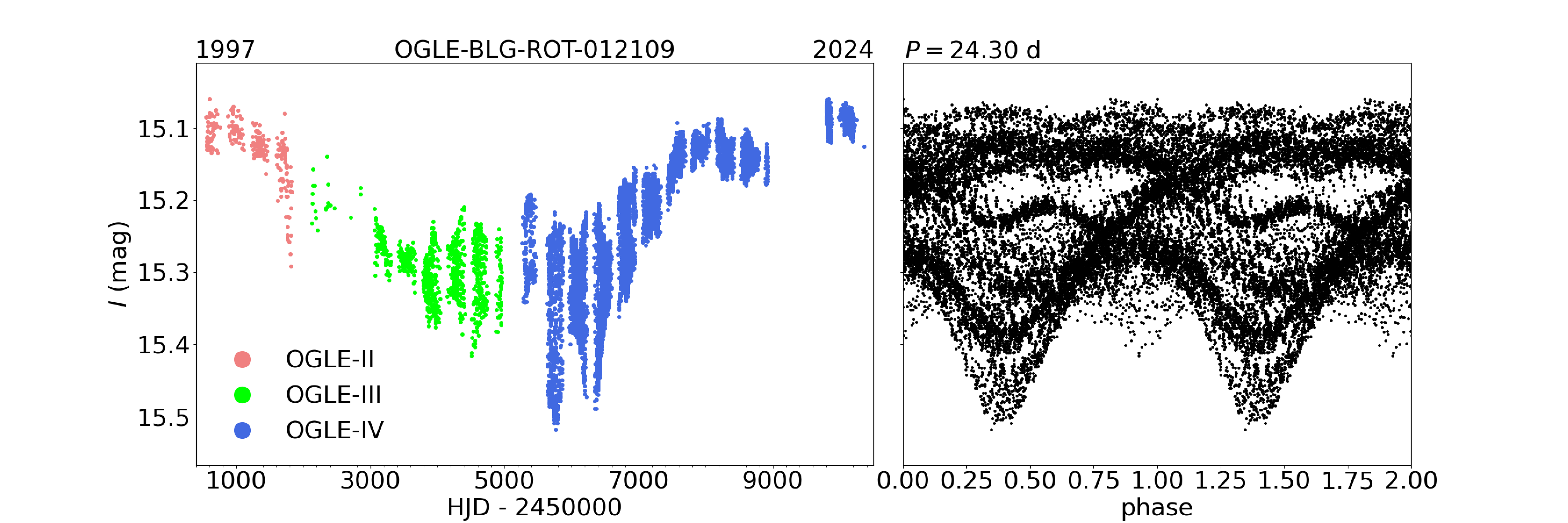}}
\vskip3pt\FigCap{Same as Fig.~4, but five other examples of rotating variables are presented.}
\end{figure}

\begin{figure}[p]
\centerline{\includegraphics[width=11.6cm]{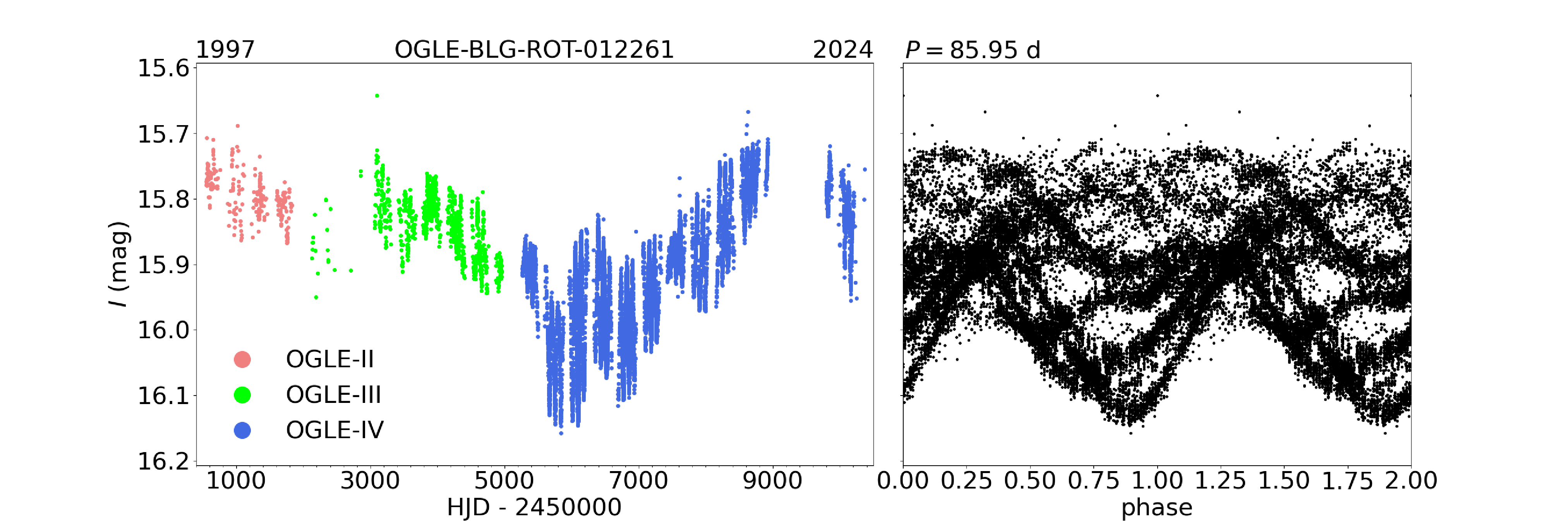}}
\centerline{\includegraphics[width=11.6cm]{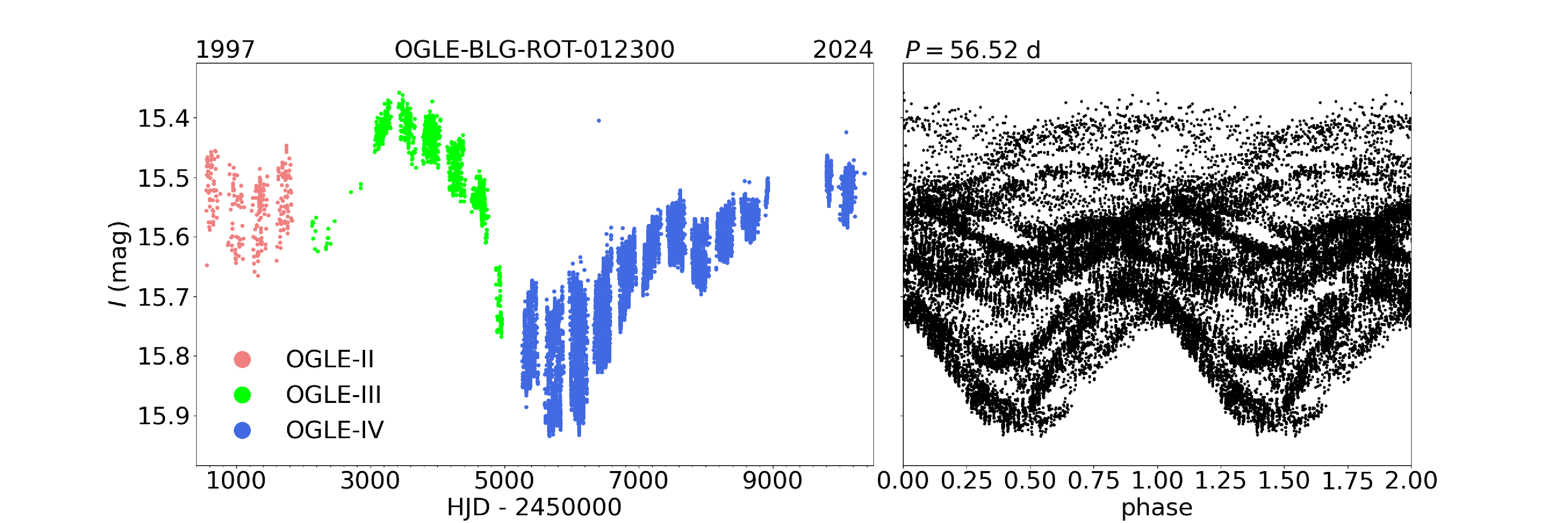}}
\centerline{\includegraphics[width=11.6cm]{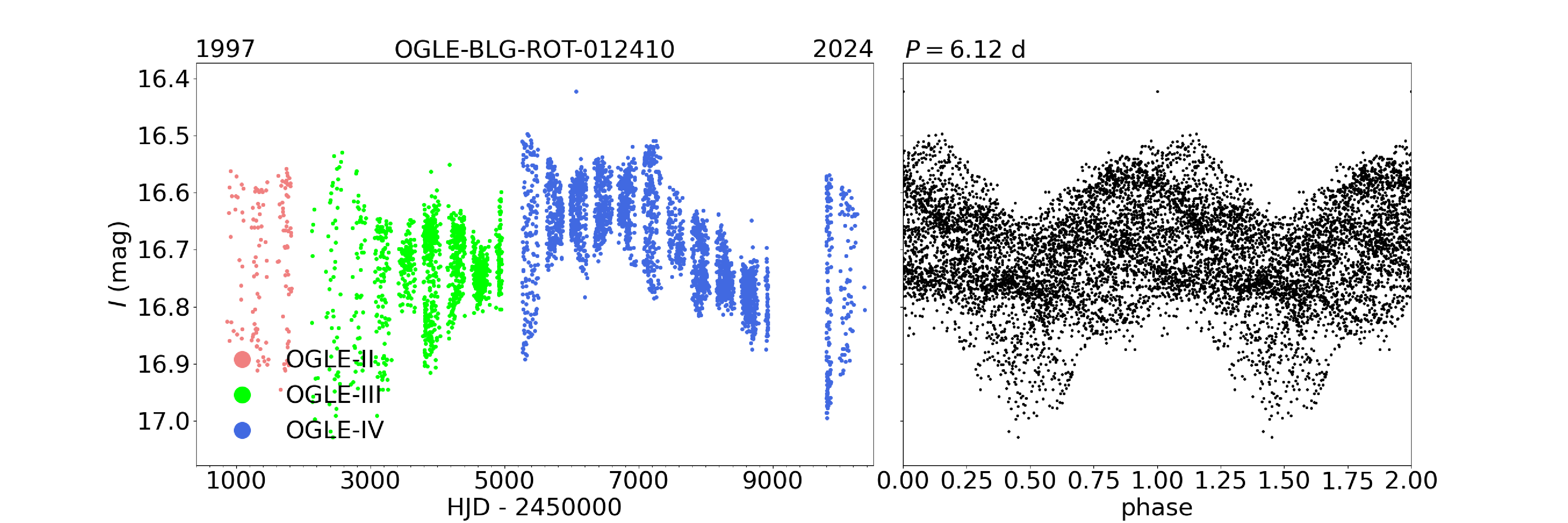}}
\centerline{\includegraphics[width=11.6cm]{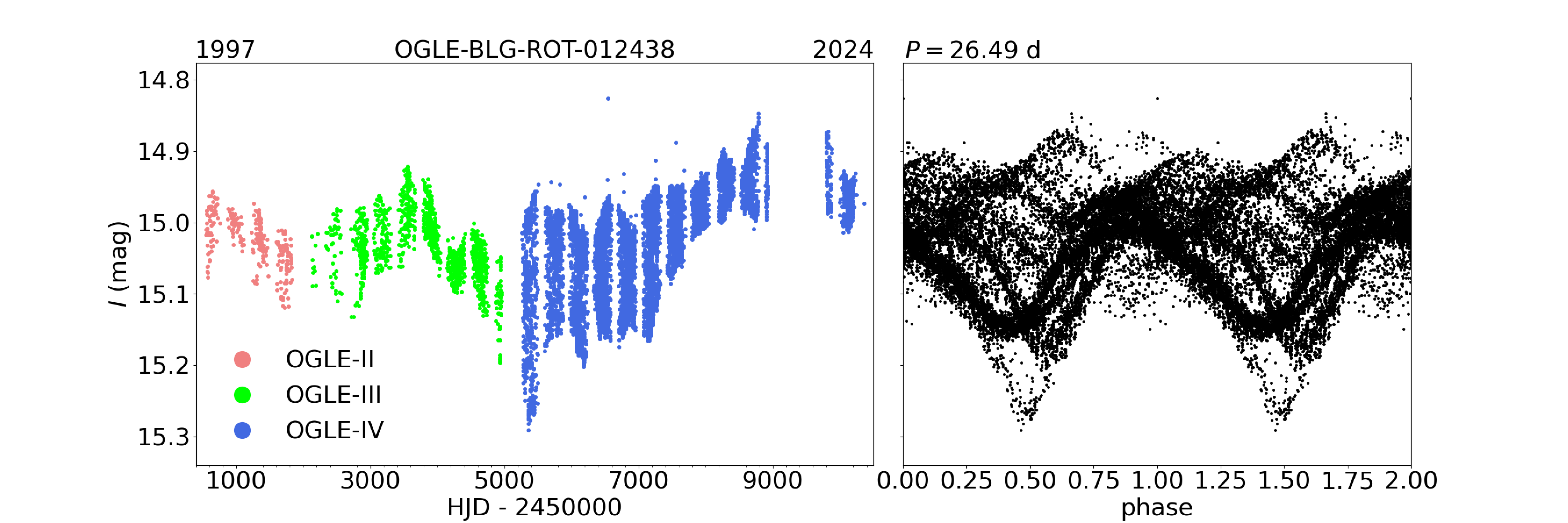}}
\centerline{\includegraphics[width=11.6cm]{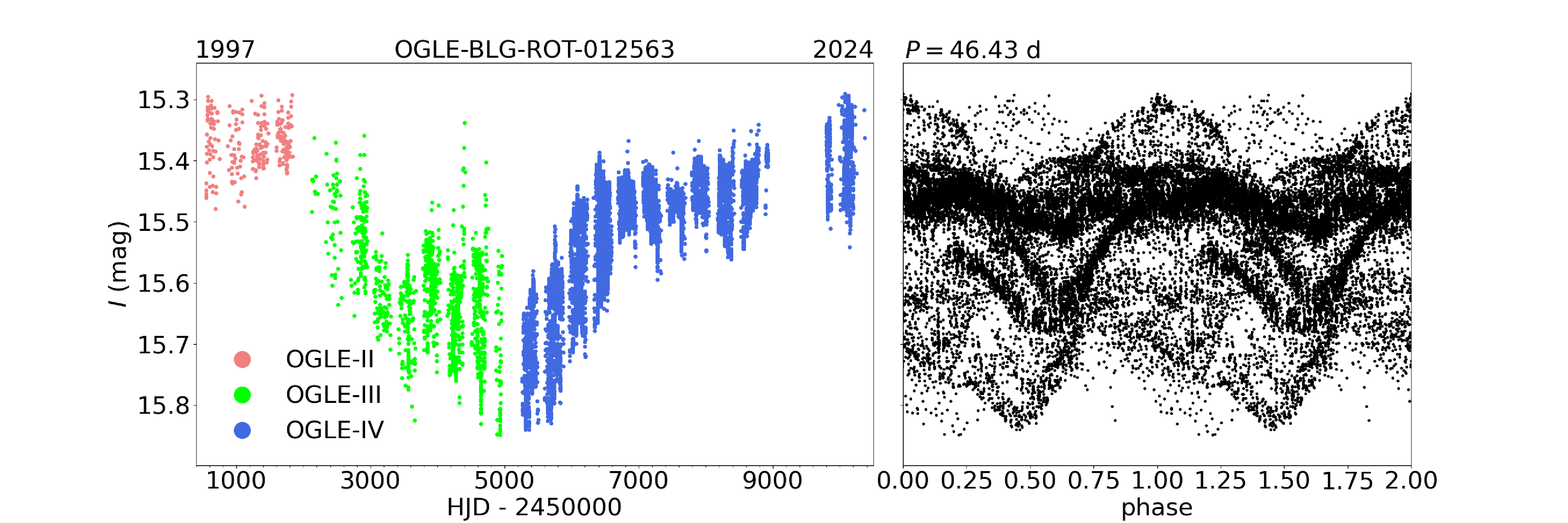}}
\vskip3pt\FigCap{Same as Fig.~4, but five other examples of rotating variables are presented.}
\end{figure}

\begin{figure}[p]
\centerline{\includegraphics[width=11.6cm]{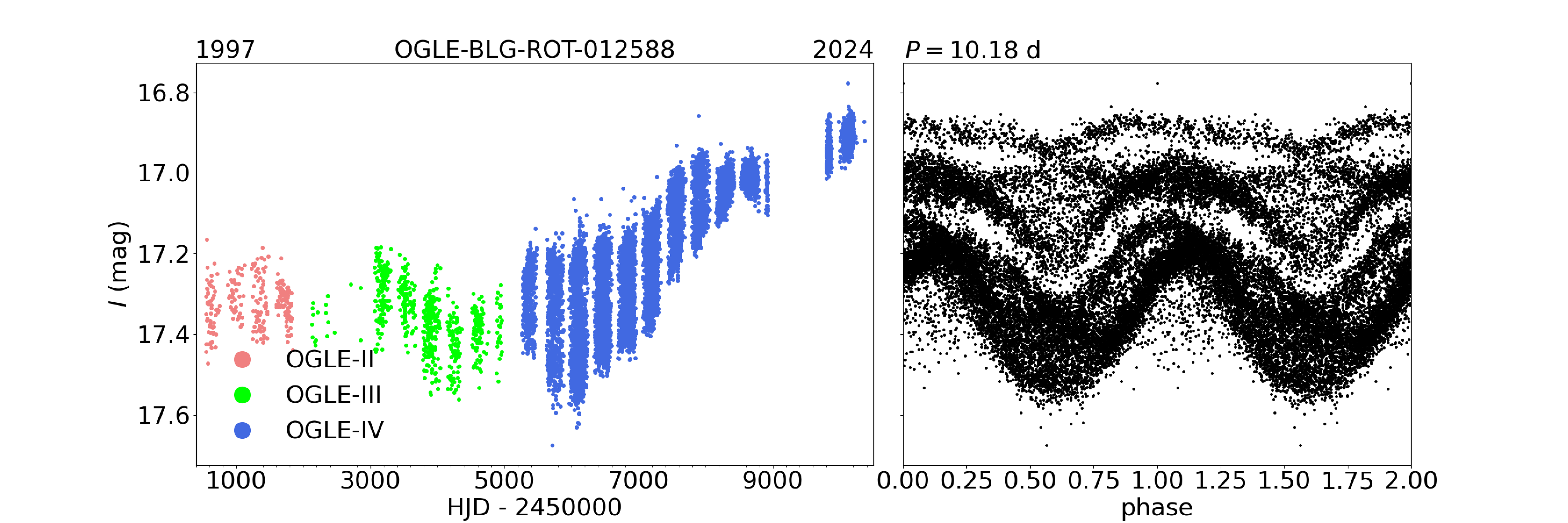}}
\centerline{\includegraphics[width=11.6cm]{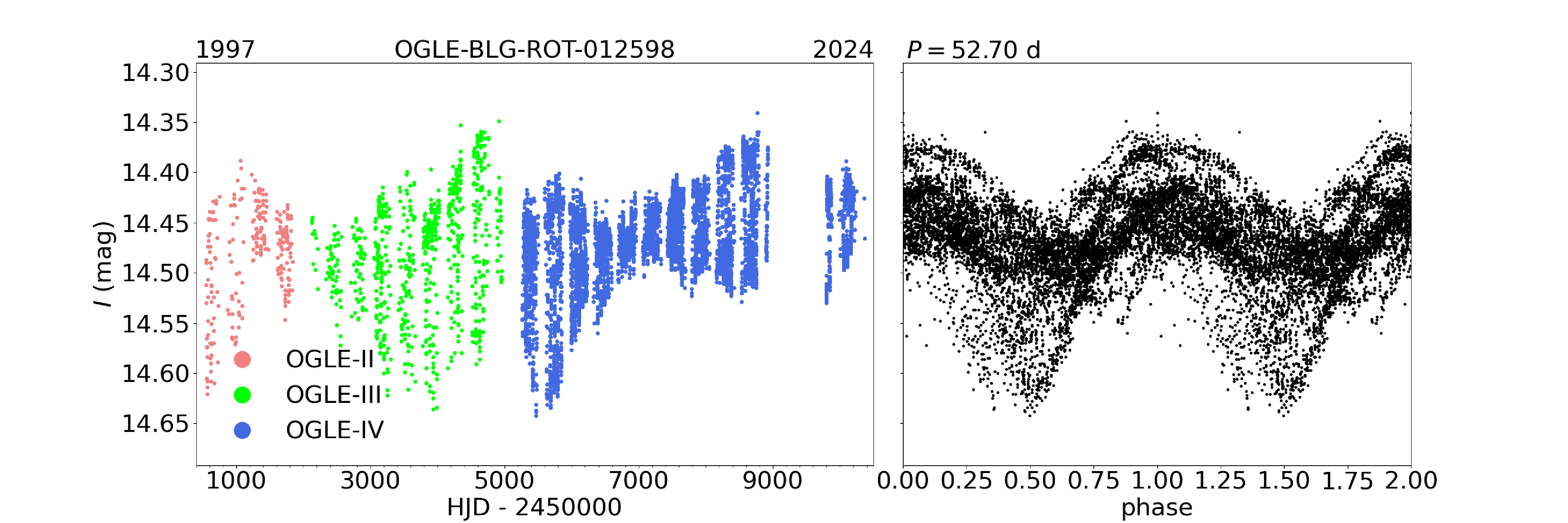}}
\centerline{\includegraphics[width=11.6cm]{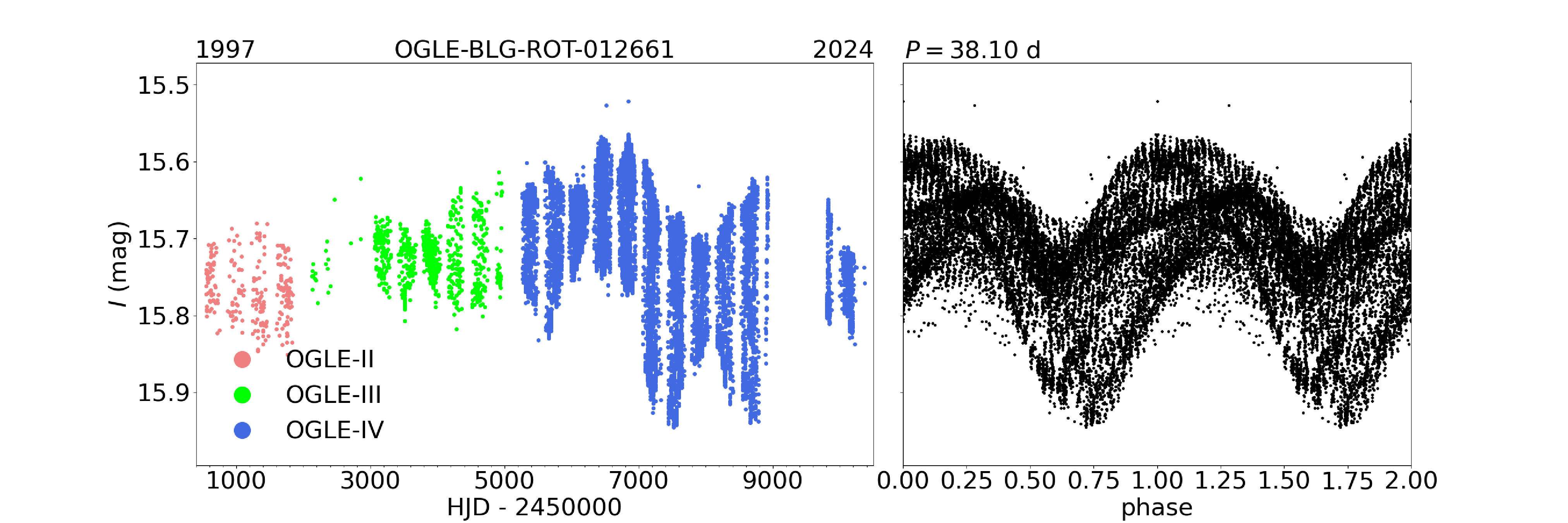}}
\centerline{\includegraphics[width=11.6cm]{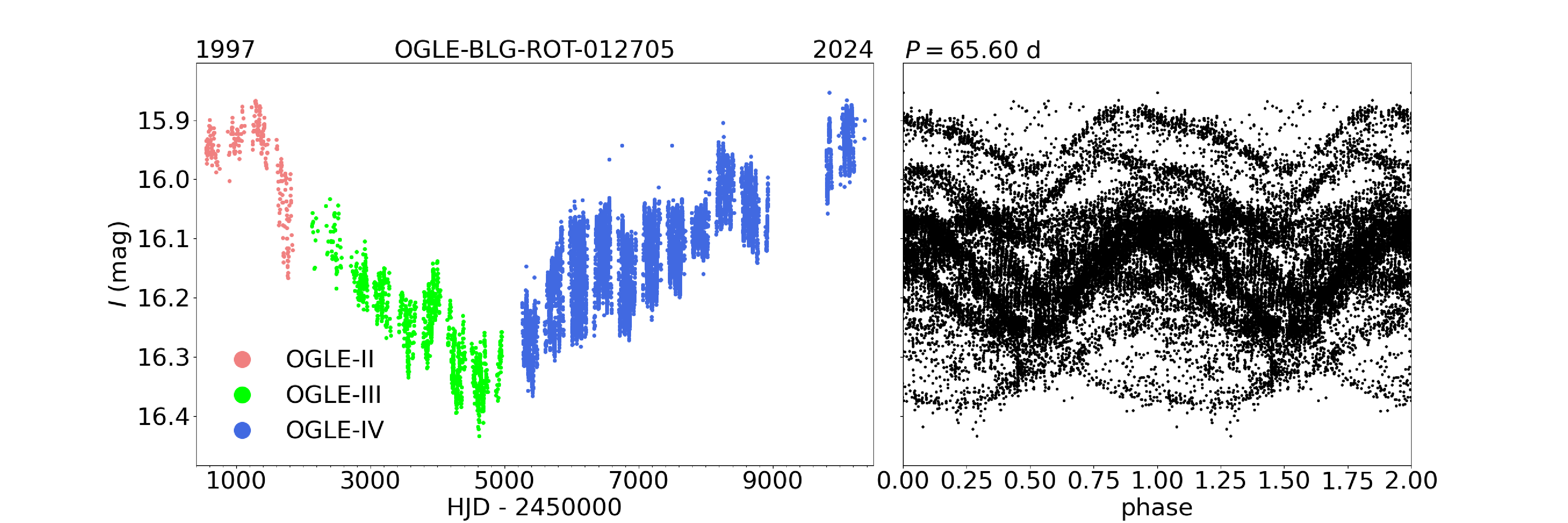}}
\centerline{\includegraphics[width=11.6cm]{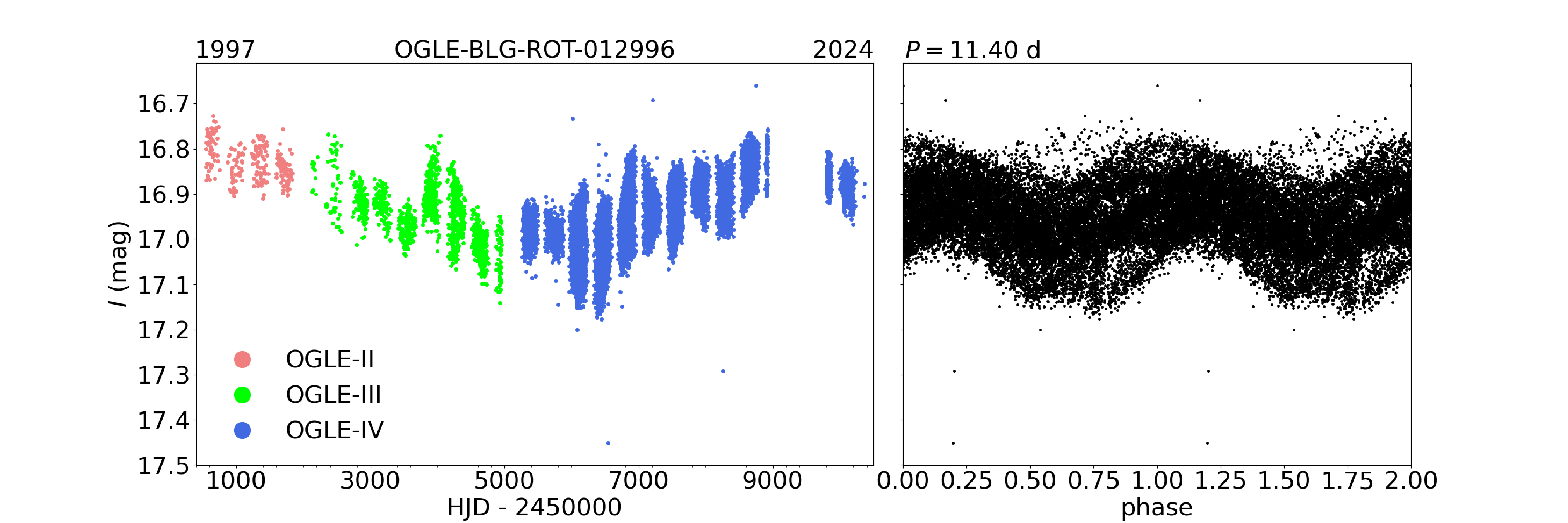}}
\vskip3pt\FigCap{Same as Fig.~4, but five other examples of rotating variables are presented.}
\end{figure}

\begin{figure}[p]
\centerline{\includegraphics[width=11.6cm]{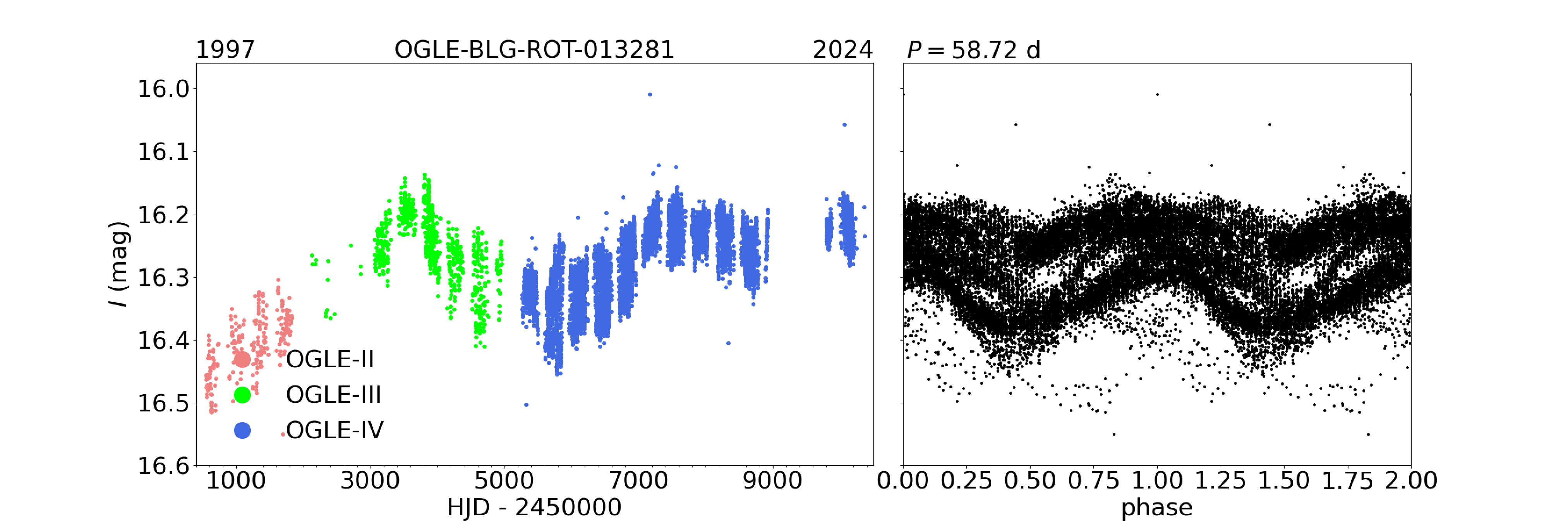}}
\centerline{\includegraphics[width=11.6cm]{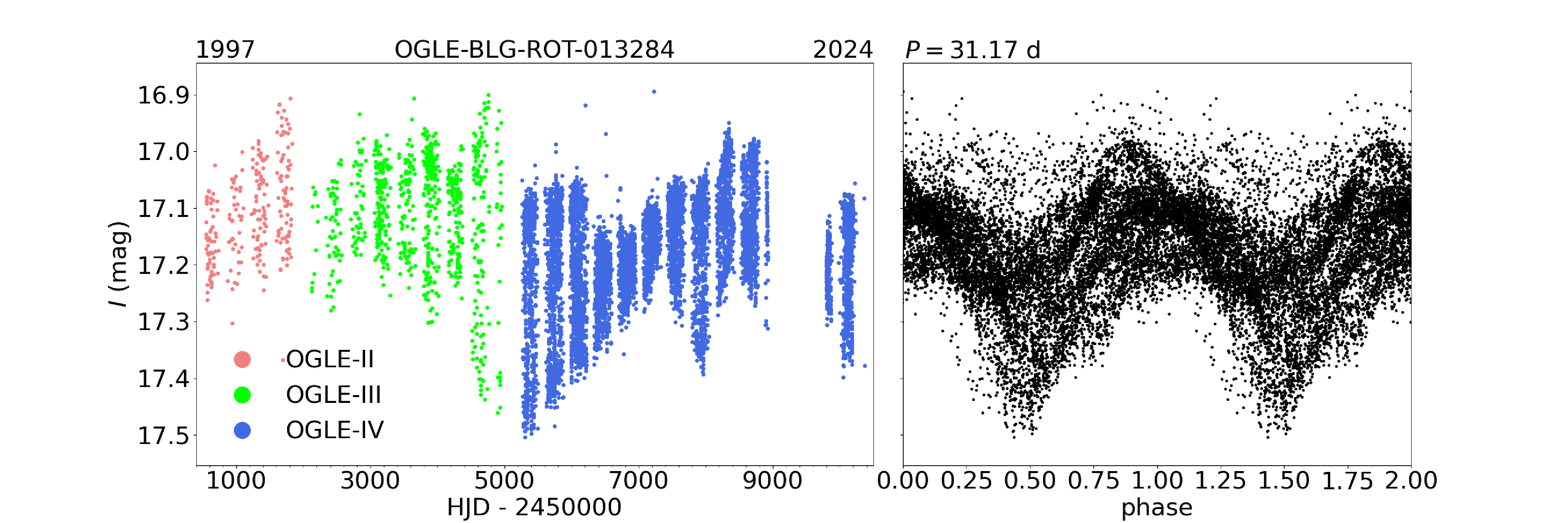}}
\centerline{\includegraphics[width=11.6cm]{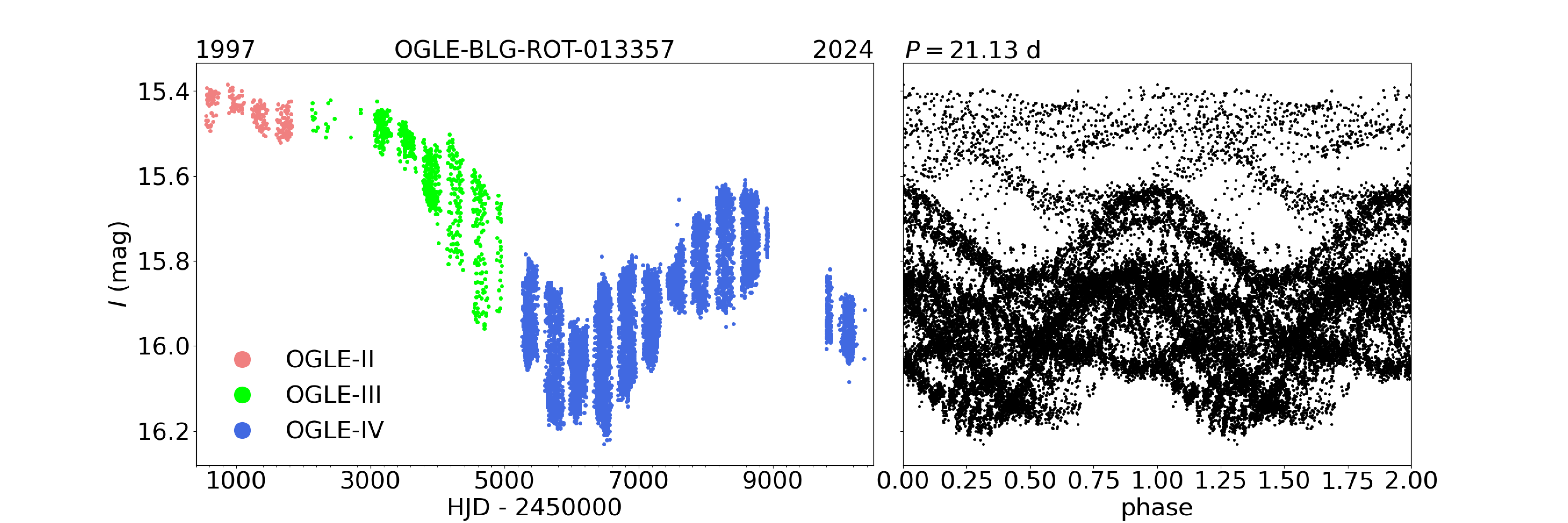}}
\centerline{\includegraphics[width=11.6cm]{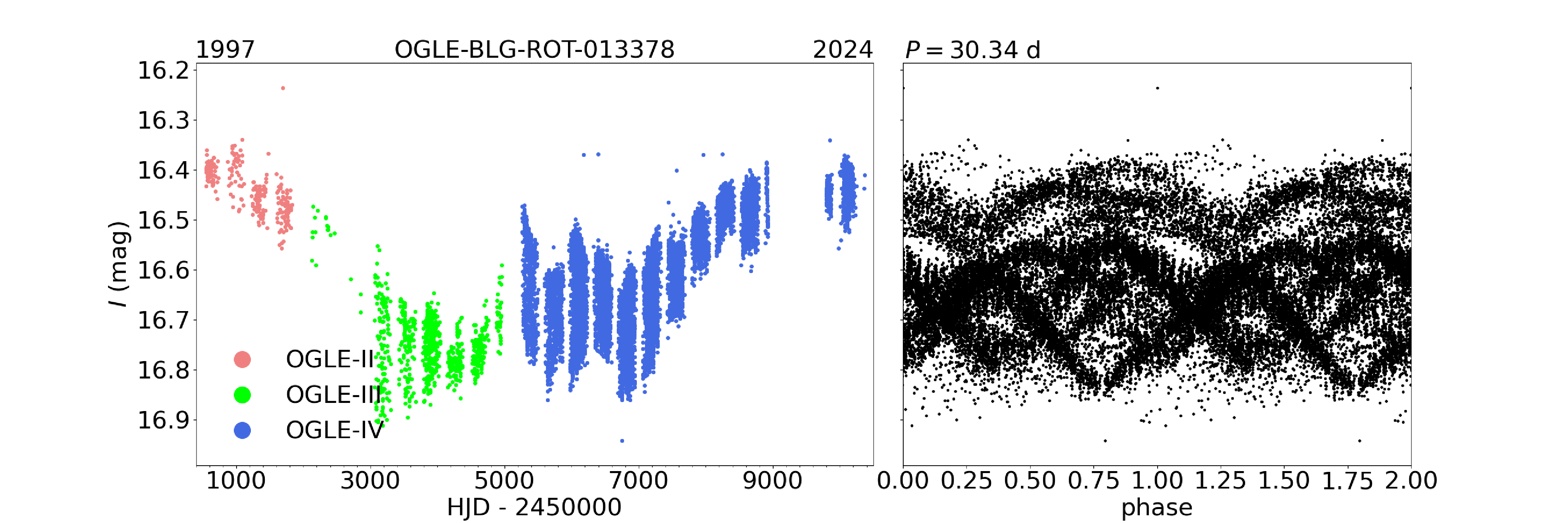}}
\centerline{\includegraphics[width=11.6cm]{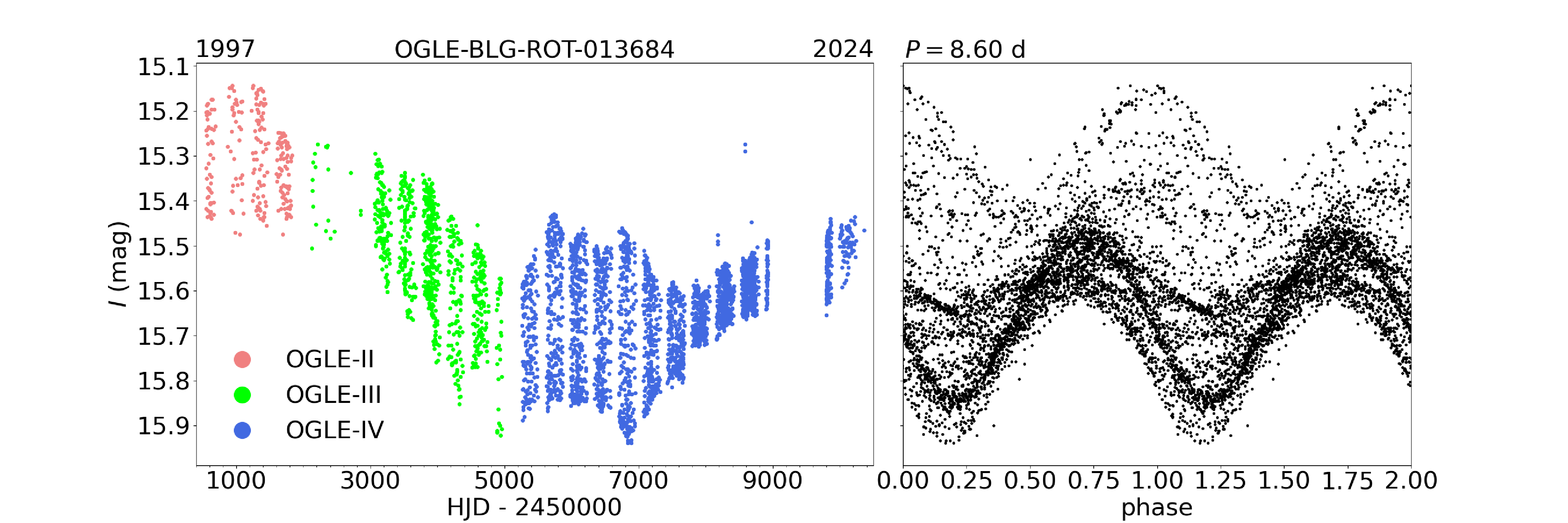}}
\vskip3pt\FigCap{Same as Fig.~4, but five other examples of rotating variables are presented.}
\end{figure}

\begin{figure}[p]
\centerline{\includegraphics[width=11.6cm]{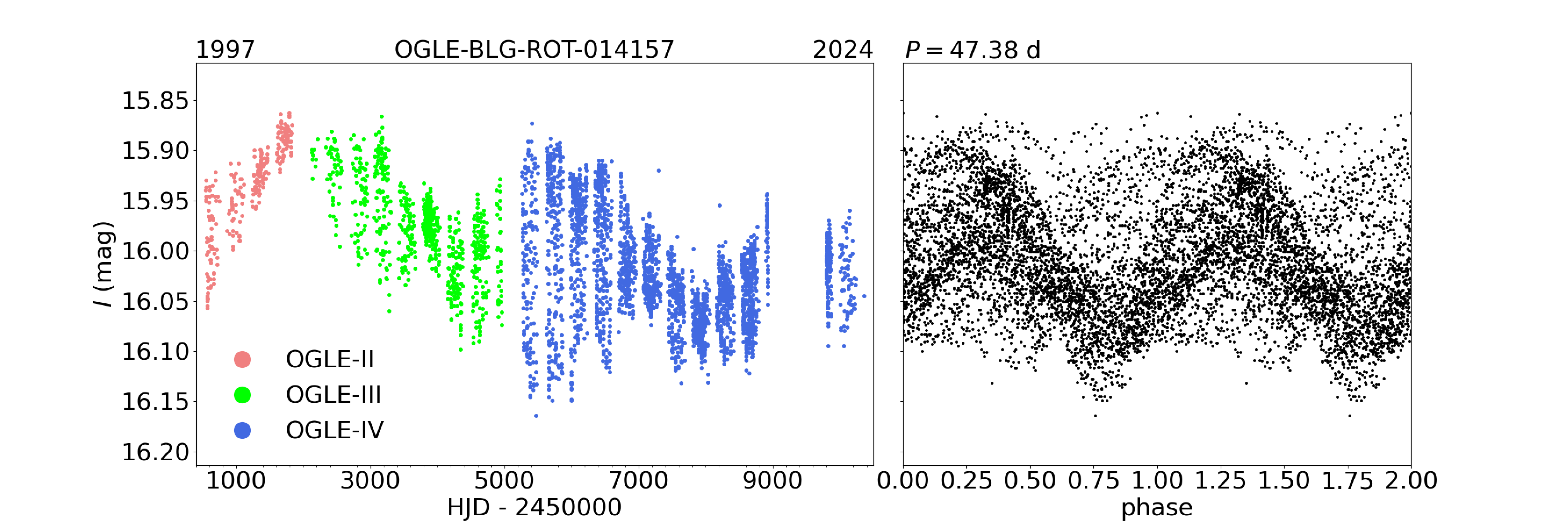}}
\centerline{\includegraphics[width=11.6cm]{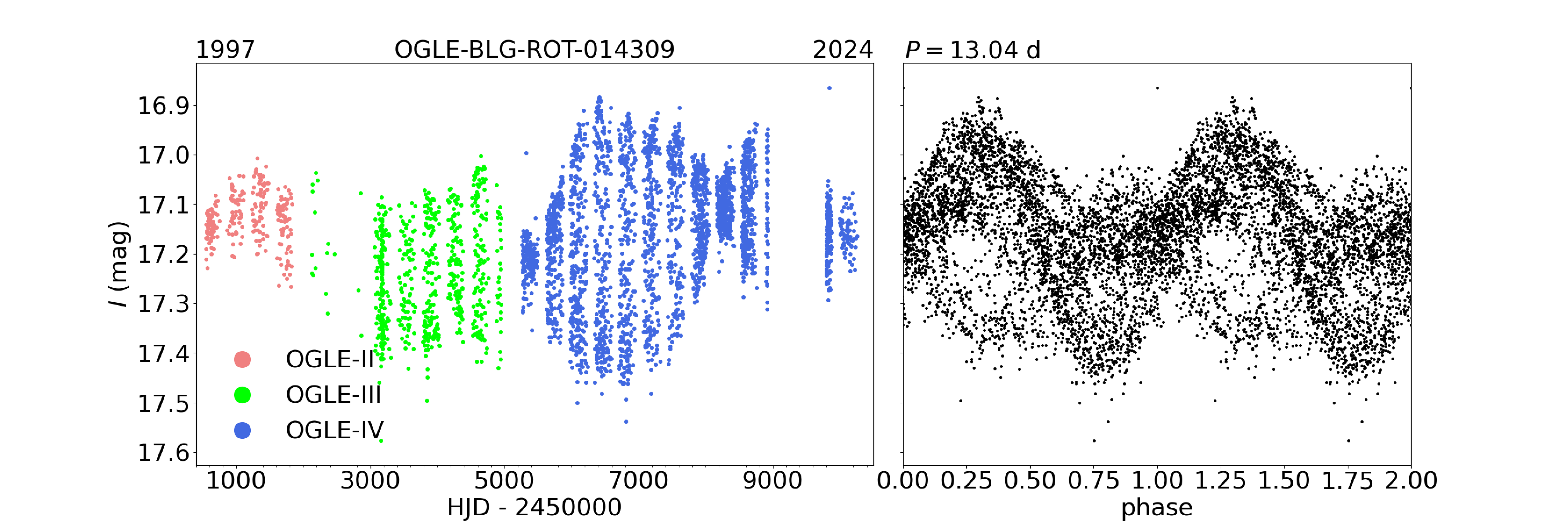}}
\centerline{\includegraphics[width=11.6cm]{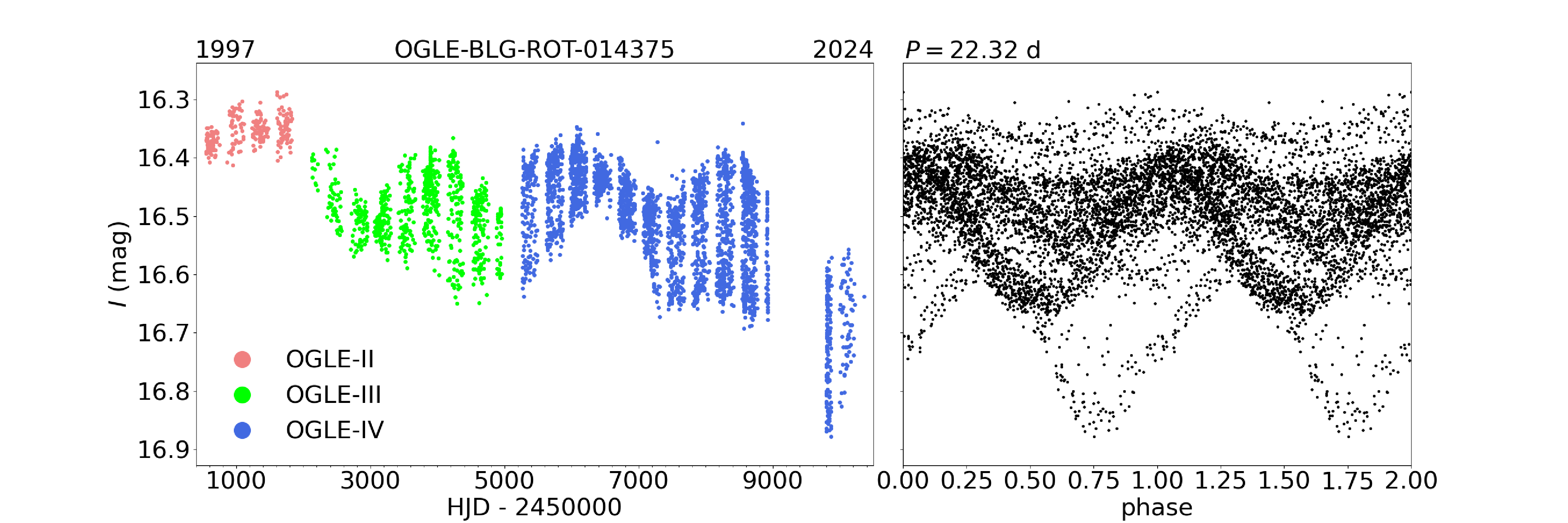}}
\centerline{\includegraphics[width=11.6cm]{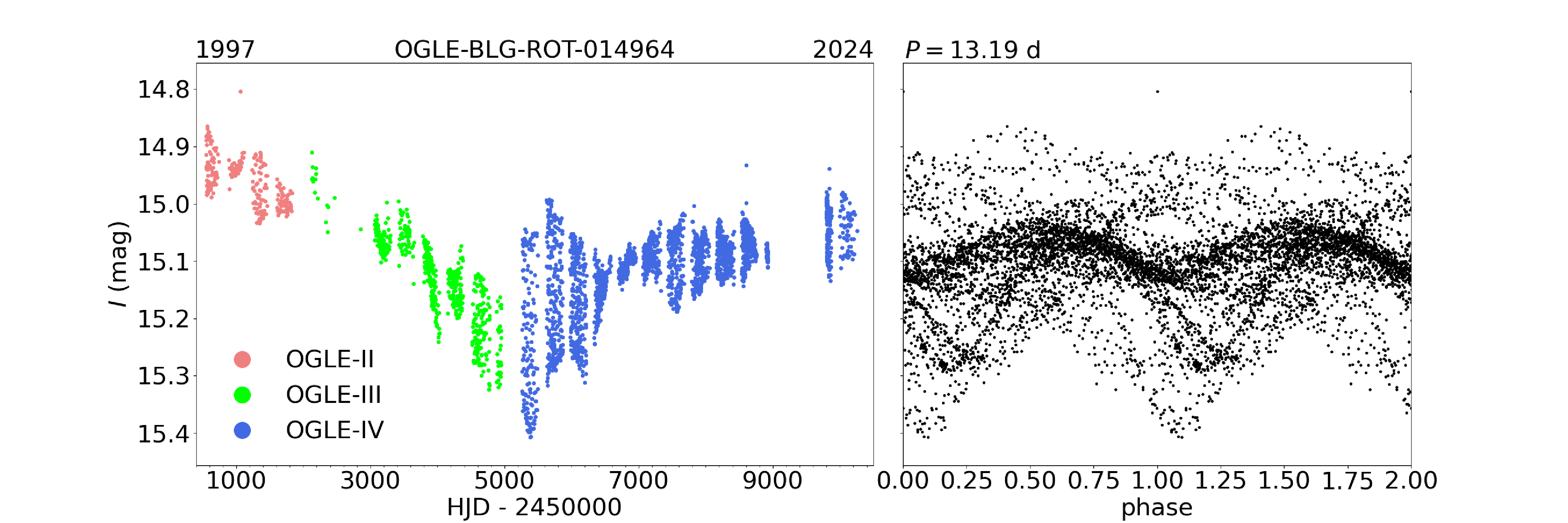}}
\centerline{\includegraphics[width=11.6cm]{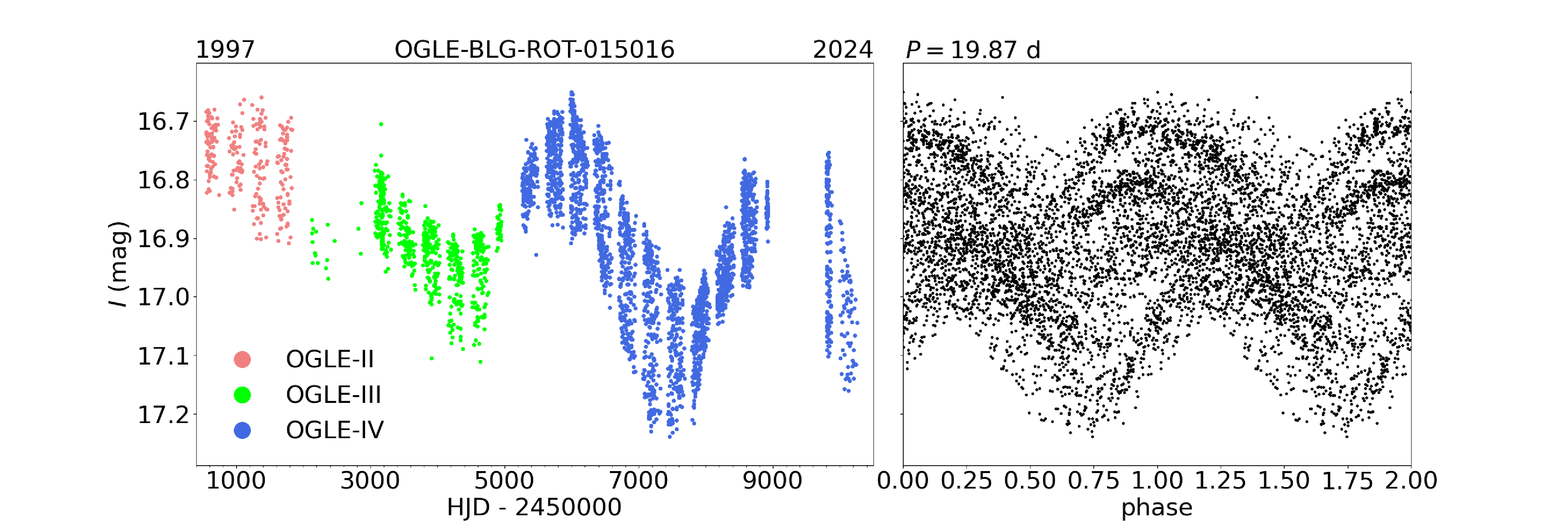}}
\vskip3pt\FigCap{Same as Fig.~4, but five other examples of rotating variables are presented.}
\end{figure}

\begin{figure}[p]
\centerline{\includegraphics[width=11.6cm]{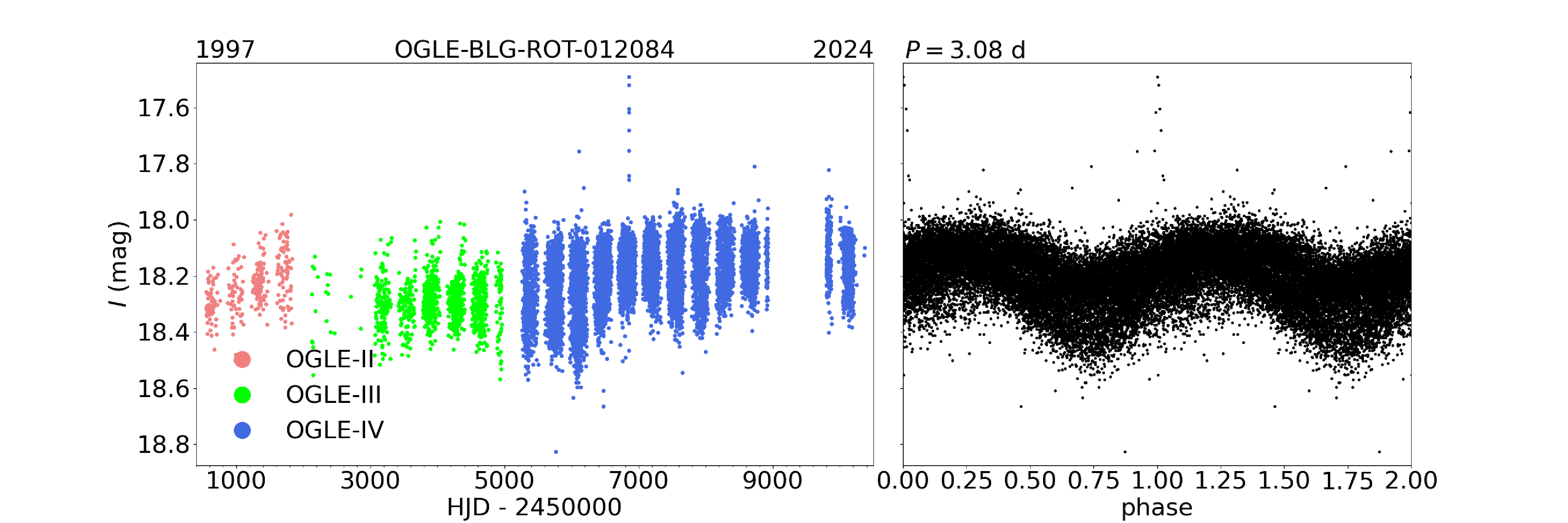}}
\centerline{\includegraphics[width=11.6cm]{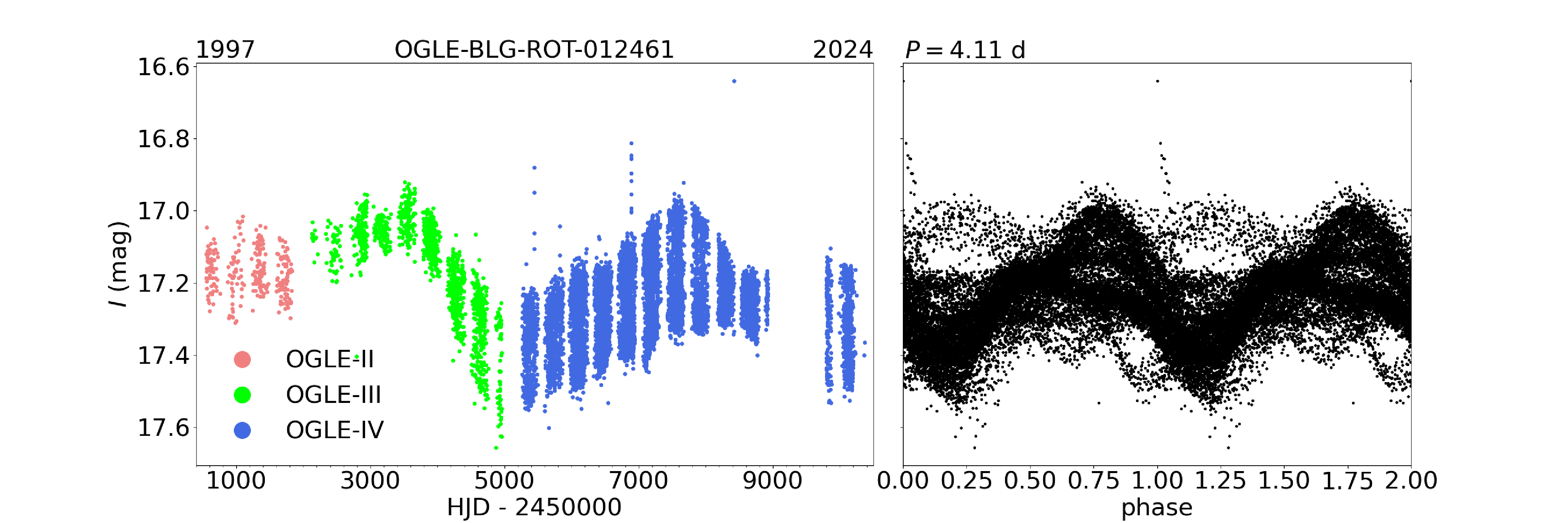}}
\centerline{\includegraphics[width=11.6cm]{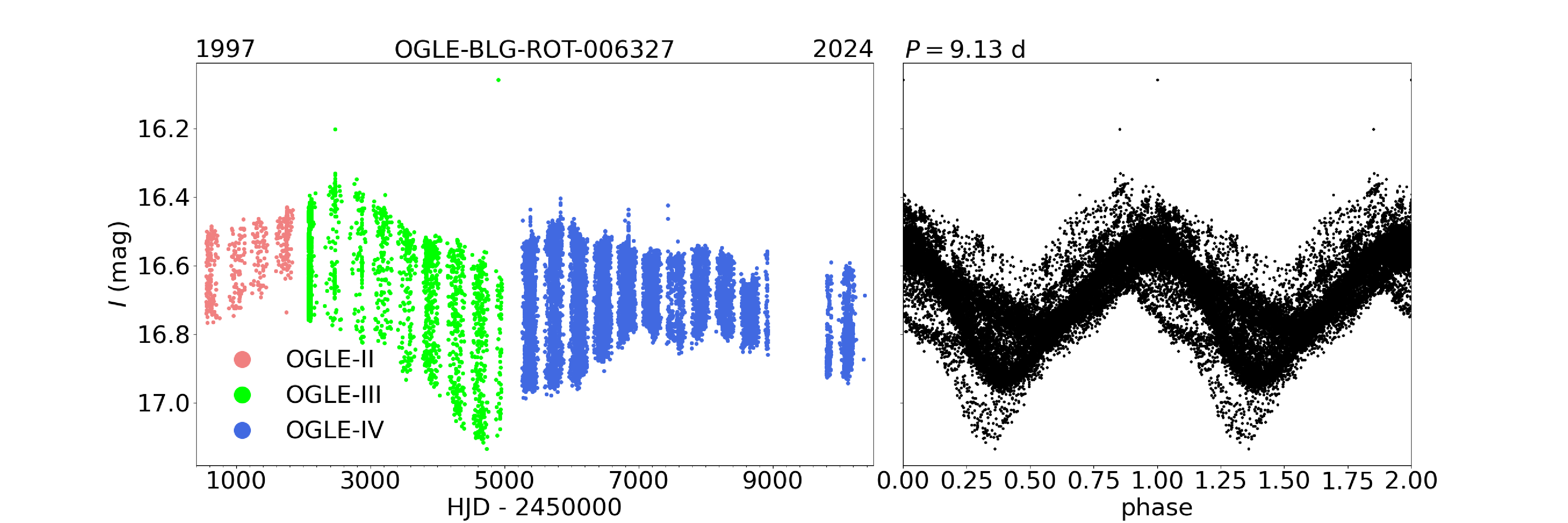}}
\centerline{\includegraphics[width=11.6cm]{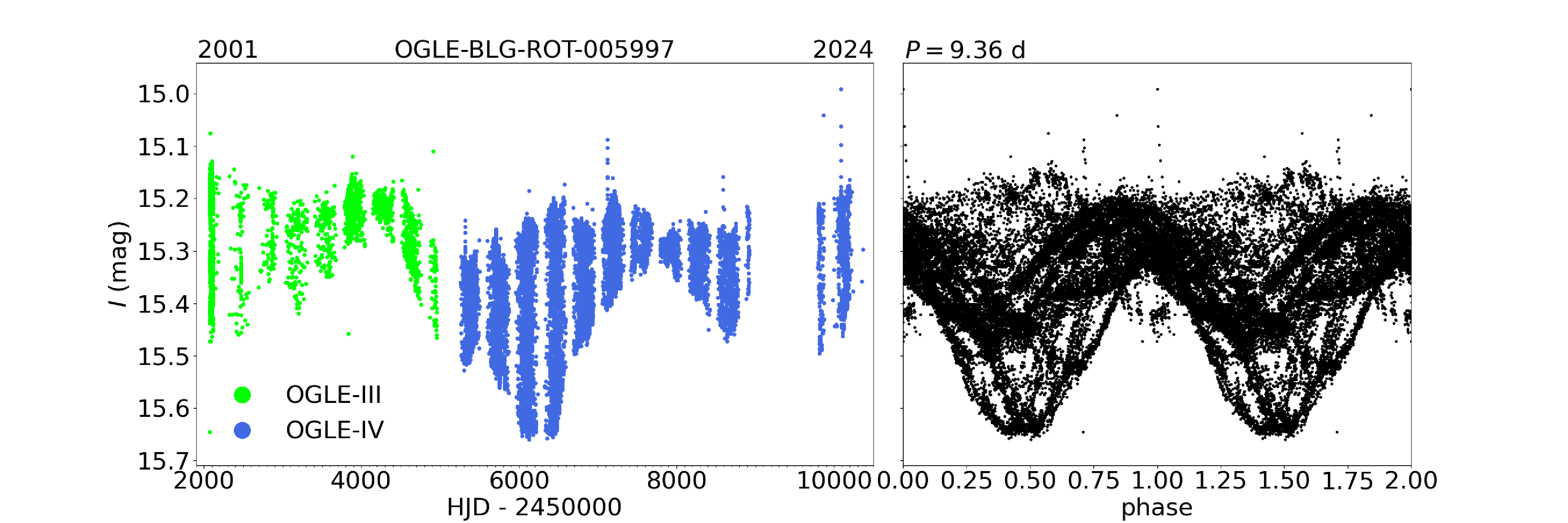}}
\centerline{\includegraphics[width=11.6cm]{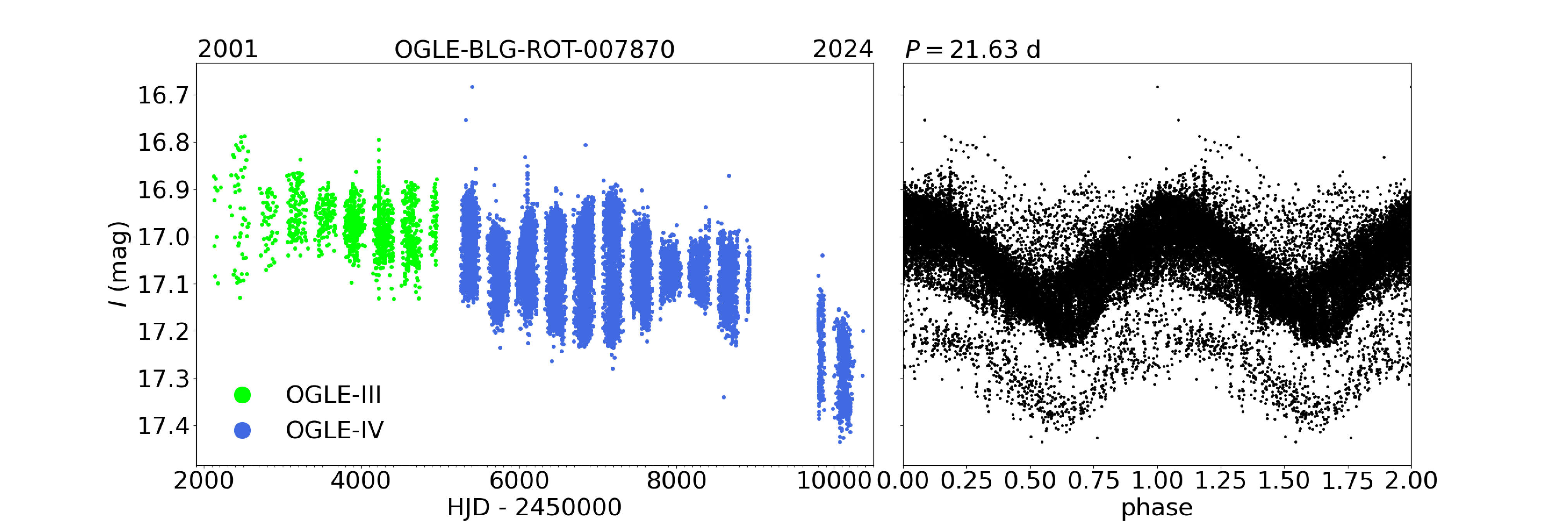}}
\vskip3pt
\FigCap{Five examples of stars in which variability we detected flares activity.}
\end{figure}

\vglue9pt
\begin{figure}[htb]
\centerline{\includegraphics[width=14cm]{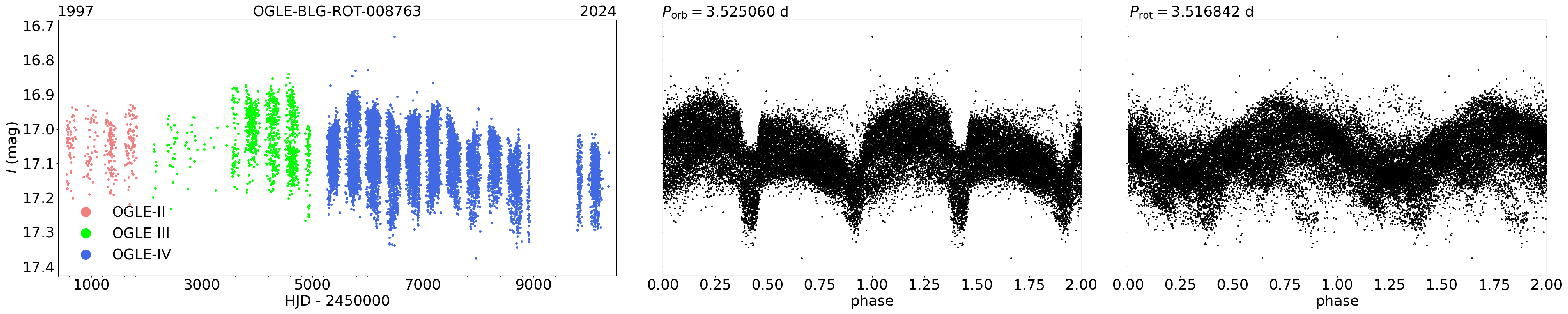}}
\centerline{\includegraphics[width=14cm]{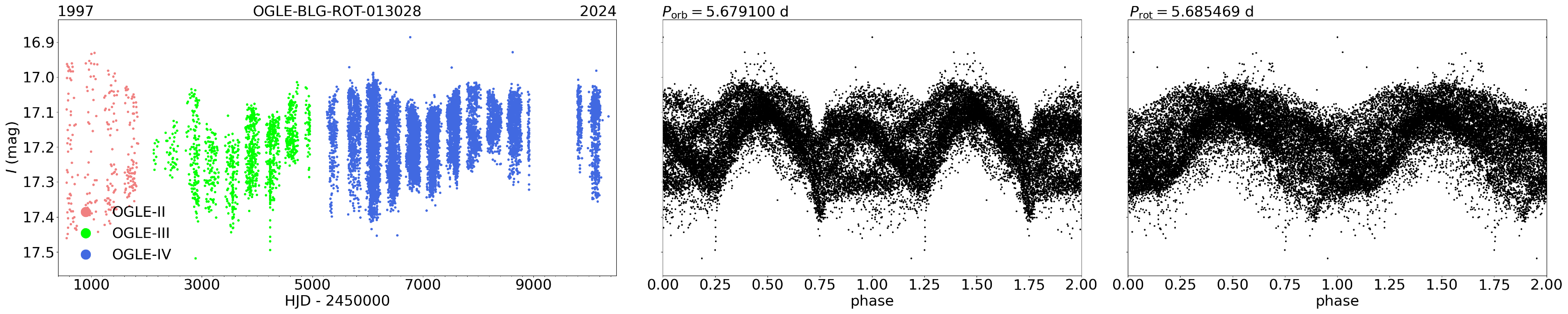}}
\centerline{\includegraphics[width=14cm]{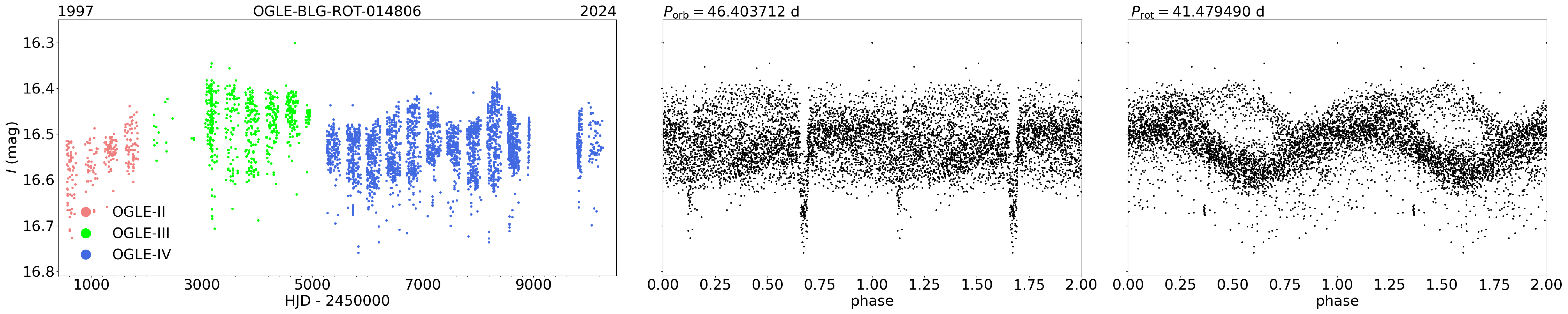}} 
\centerline{\includegraphics[width=14cm]{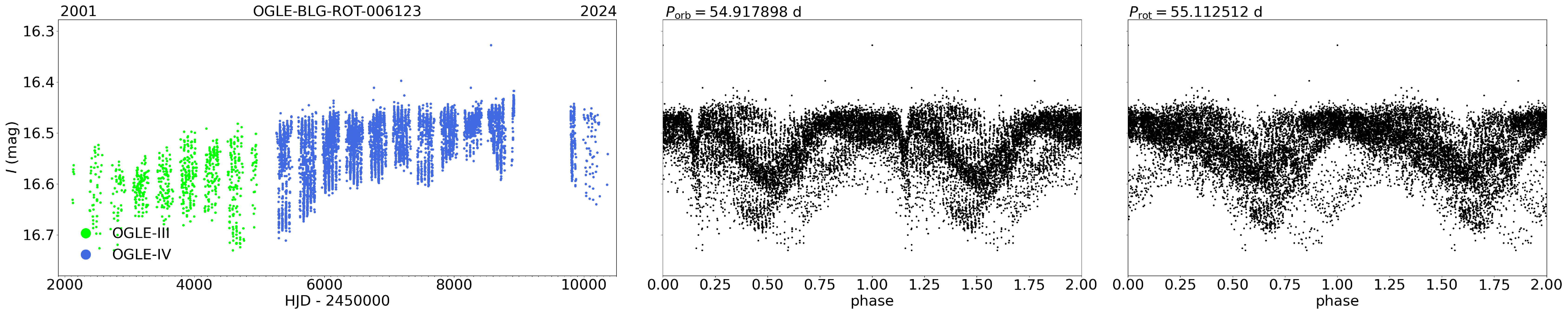}}
\centerline{\includegraphics[width=14cm]{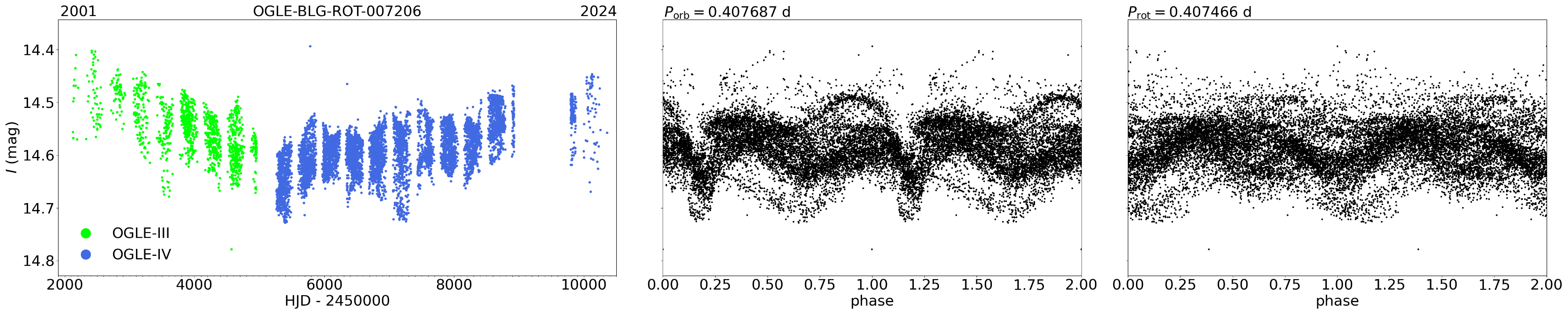}}
\vskip3pt\FigCap{Five examples of RS CVn binaries from our collection.}
\end{figure}

\subsection{Flaring Stars and Eclipsing Binaries with Spots}
Flares are another indicator of stellar magnetic activity, following
stellar spots. Classical flares on cool stars are phenomena of the sudden
release of the energy stored by the magnetic field in active regions of
stars through magnetic reconnection. Recent studies show that flares also
occur on hot stars (Balona 2012, Bai and Esamdin 2020), but the mechanism
producing flares is slightly different compared to cool stars and is
related to shocks and radiatively driven winds. In general, flare activity
depends on the strength of the magnetic field, which can be stronger with
higher rotational velocity. Flares are usually observed as asymmetric
brightenings in the light curves (rapid rise and slow decline). However,
releases of energy observed as flares can also be observed as symmetric
rise and decline (Mr\'oz 2023).  An analysis of the statistical properties of
flares discovered in the OGLE data can be found in Iwanek \etal (2019). In
Fig.~11, we present five objects from our collection in which we detected
flare activity.

RS CVn-type stars are binaries consisting of a hot primary component, which
is a subgiant or giant of spectral type F-K, and a cooler secondary
component, usually a dwarf or a subgiant of type G-M (Biermann and Hall
1976). RS Canum Venaticorum is the prototype of the whole class, and its
light variations were first observed by Hoffmeister (1915).  The nature of
the RS CVn objects are very complex and is caused by eclipses (in case of
the high inclination), ellipticity, but also spots. In the system, at least
one component (usually the primary one) is heavily spotted (with spots
covering even 10\% of the stars' surface, Vogt \etal 1999). RS~CVn stars
also show flare activity (Pietrukowicz \etal 2013, Kunt and Dal 2017,
Lahiri \etal 2023), and activity cycles (Mart{\'i}nez \etal 2022). The stars in
the system are usually tidally synchronized, meaning that the active
primary component rotates in exact synchronization with the orbital motion
(\ie the orbital period is equal to the average rotation period of the
primary component). However, due to differential rotation and the evolution
of starspots on the stellar surface, the period of the observed starspot
wave can be slightly different from the orbital period. Yet, it is
estimated that about 15\% of RS CVn stars do not rotate synchronously or
even nearly synchronously (Hall and Henry 1990), and the ratio of the
rotation period ($P=P_{\rm rot}$) to the orbital period ($P_{\rm orb})$ can
reach even a factor of several. Examples include two extreme systems:
$\lambda$ And with the ratio equal to 2.6 and V4138~Sgr with the ratio
equal to 4.6 (Hall 1992).

In Fig.~12, we present five examples of RS CVn-type stars from our
collection. In this figure, we show phase-folded light curves with orbital
and rotation periods.  In the left panel, we show unfolded light curves,
while in the middle and right panels, we show phase-folded light curves
with orbital period $P_{\rm orb}$ and rotation period $P_{\rm rot}$,
respectively. The periods are indicated above the phase-folded plots. The
dates at the tops of the unfolded light curves mark the year when the
observations started and the year of the last used observations in the
collection. Different OGLE phases are marked with different colors (red,
green, and blue).

\Section{Conclusions}
We presented the collection of over 18\,000 rotating variables with clear
magnetic field manifestation, \ie spots or flares. With this paper, we
provide long-term, time-series photometry in the Cousins {\it I}- and
Johnson {\it V}-band obtained from the second phase to the current fourth
phase of the OGLE project (since 1997), finding charts, and basic
observational parameters, \ie equatorial coordinates, rotation periods,
mean brightness and amplitudes in both bands. The meticulous classification
process, involving semi-automatic methods and visual inspection, ensures
the high purity of the collection. This collection provides an invaluable
dataset laying the groundwork for future studies of magnetic activity,
including activity cycles.

\Acknow{This work has been supported by the Polish National Science
  Centre grant PRELUDIUM 18 No. 2019/35/N/ST9/02474 to P.I. I.S. has been
  supported by the National Science Center, Poland, grant
  No. 2022/45/B/ST9/00243. For the purpose of Open Access, the author has
  applied a CC-BY public copyright license to any Author Accepted
  Manuscript (AAM) version arising from this submission. This research has
  made use of the International Variable Star Index (VSX) database,
  operated at AAVSO, Cambridge, Massachusetts, USA. We would like to thank
  OGLE observers for their contribution to the collection of the
  photometric data used in this paper.}

\end{document}